\documentclass[11pt]{article}

\usepackage{amsmath,amssymb}
\usepackage{amsthm}
\usepackage{graphicx,psfrag,epsf,epsfig}
\usepackage{enumerate}
\usepackage{verbatim}
\usepackage{color, colortbl}
\usepackage{setspace}

\usepackage{natbib}
 \bibpunct[, ]{(}{)}{,}{a}{}{,}%
 \def\newblock{\ }%
\usepackage{use-with-onehalfS}

\newcommand{\boldx}{\mbox{$\mathbf{x}$}}

\newcommand{\boldphi}{\mbox{$\mathbf{\varphi}$}}

\def\minimize{\mathop{\rm minimize}\limits}
\def\argmin{\mathop{\rm argmin}\limits}

\newtheorem{definition}{Definition}

\begin{document}



\title{\bf{A Multi-objective Exploratory Procedure for Regression Model Selection}}
 
\author{Ankur Sinha, Pekka Malo and Timo Kuosmanen\\
Department of Information and Service Economy\\ Aalto University School of Business\\
       PO Box 1210, FIN-00101, Helsinki, Finland\\ 
\texttt{Firstname.Lastname@aalto.fi}}
\date{}
\maketitle

\begin{center}
\textbf{Abstract}
\end{center}

Variable selection is recognized as one of the most critical steps in statistical modeling. The problems encountered in engineering and social sciences are commonly characterized by over-abundance of explanatory variables, non-linearities and unknown interdependencies between the regressors. An added difficulty is that the analysts may have little or no prior knowledge on the relative importance of the variables. To provide a robust method for model selection, this paper introduces the Multi-objective Genetic Algorithm for Variable Selection (MOGA-VS) that provides the user with an optimal set of regression models for a given data-set. The algorithm considers the regression problem as a two objective task, and explores the Pareto-optimal (best subset) models by preferring those models over the other which have less number of regression coefficients and better goodness of fit. The model exploration can be performed based on in-sample or generalization error minimization. The model selection is proposed to be performed in two steps. First, we generate the frontier of Pareto-optimal regression models by eliminating the dominated models without any user intervention. Second, a decision making process is executed which allows the user to choose the most preferred model using visualizations and simple metrics. The method has been evaluated on a recently published real dataset on Communities and Crime within United States.


\section{Introduction}
Model selection tasks commonly appear in many branches of science. Investigators are often interested in finding the best model for the dependent variable which leads to a good quality of fit and parsimony. A compromise is to be made between fitness and parsimony, as inclusion of too many dependent variables lead to loss in precision of the regression coefficients and omitting important factors lead to a mis-estimation of the regression coefficients and biased prediction (\cite{ms-murtaugh}). This trade-off makes the model selection task a two objective problem. 
However, most of the existing approaches have handled the model selection task as a single objective problem by using various penalized model selection criteria (such as AIC and BIC); see e.g. \cite{ms-jeffreys, miller02, ms-burnham, ms-mackay, ms-gregory, zhu06} and references therein. Despite a lot of work in the direction of model selection techniques, there is no single method which can be utilized for all problems. This is explained by the fact that model selection task is inherently not a single objective problem with a uniquely defined solution. Instead, each selection criterion or single objective method is bound to produce different results, because they work by giving higher or lower importance to either fitness or parsimony. In the past years, ideas from the field of computer science and machine learning have been applied to the field of statistics, particularly for problems with large number of independent variables. Some of the examples are random forests (\cite{breiman01}), support vector machines (\cite{vapnik95}), and boosting (\cite{freund96,hofner11}). In this paper, we utilize the principles from the field of evolutionary computation, to handle the model selection problem through a bi-objective approach.

We propose a multi-objective genetic algorithm for variable selection (MOGA-VS), which draws insights from the advances in the field of evolutionary computation (\cite{deb-book-01, carlos-book}). In MOGA-VS, the model selection task is considered as a multi-objective optimization problem, where the first objective is to reduce the complexity of the model (or reduce the number of coefficients) and the second objective is to maximize the goodness-of-fit (or minimize mean squared error).  By doing so, the suggested approach differs from the existing methods in two important ways. First, instead of attempting to arrive at a single model candidate, the method produces a collection of Pareto-optimal\footnote{The notion of Pareto-optimality is synonymous to the best-subset.} regression models from which the most preferred model can be chosen. The second difference follows from the separation of optimization process from choosing a particular trade-off between goodness-of-fit and model parsimony. The problem of finding all optimal trade-offs is performed without any user-intervention, whereas the task of selecting an optimal balance between the two objectives is best left as a user's preference-based decision. The proposed algorithm can also be viewed as a method for exploring the best-subset, which is a tedious task especially for large scale problems. The algorithm can also be easily extended to minimize the generalization error in order to obtain a Pareto-optimal frontier of models based on their generalization performance. The multi-objective optimization task using MOGA-VS is followed by the decision making process, which may be performed using a combination of visual tools and metrics.

The rest of this paper is organized as follows. Section~\ref{sec:mop} provides a summary of the central definitions and an overview of the model selection problem within multi-objective framework. 
Section~\ref{sec:overview} gives a literature review on commonly applied model selection methods and discusses their differences to multi-objective optimization framework. The proposed MOGA-VS algorithm is presented in Section~\ref{sec:moga}. Section~\ref{sec:results} presents the results from experiments with three different datasets. One of the datasets is a recently published real dataset on Communities and Crime within United States. Comparisons with respect to well known variable selection techniques are included in the study. Section~\ref{sec:genError} extends the MOGA-VS algorithm for generalization error minimization, and provides results on one of the datasets. Finally, we provide the conclusions in Section~\ref{sec:conclusions}.

\section{Model Selection as a Multi-objective Problem}\label{sec:mop}

In this section, we formulate the regression problem as a multi-objective task, and introduce the main concepts and the notation used in multi-objective problems.


\subsection{Trade-off between complexity and fit}

The regression modeling task can be viewed as a special example of supervised learning. Let $\mathcal{Y}$ be the output space and let $\mathcal{X}=\prod_{i=1}^p\mathcal{X}_i$ denote the input space, where $\mathcal{X}_i$ is the domain of the $i$-th explanatory variable and $p$ is the total number of variables. Given a collection of data $(\mathbf{X}_j,Y_j)\in\mathcal{X}\times\mathcal{Y}$, $j=1,\dots,n$, with an unknown probability distribution $\mathcal{D}$, the purpose is to find a model $f:\mathcal{X}\to\mathcal{Y}$ with minimal error on the training set with respect to $\mathcal{D}$. To restrict the search space, the model is assumed to belong to a pre-defined {\it hypothesis space} $\mathcal{H}$. For linear regression, the hypothesis space can be written as the set of all linear functions that can be formed using some subset of the variables contained in the input space,
\begin{equation}
\mathcal{H}=\left\{x \to \sum_{k\in J}\beta_kx_k \ | \ J\subset \{1,\dots,p\}, x_k\in\mathcal{X}_k\right\}.
\end{equation}

The model selection problem follows from the fact that the hypothesis space consists of models with varying complexity. In the case of regression modeling, the hypothesis space $\mathcal{H}$ forms a nested structure $\mathcal{H}_1\subset\mathcal{H}_2\subset\cdots\mathcal{H}_d\subset\cdots\subset\mathcal{H}$, where $\mathcal{H}_d$ represents the subset of models with $d$ variables. This means that in order to find a preferred model, we need to choose the size of the hypothesis space that provides a good balance between the complexity and fit. It is well known that the larger the hypothesis space (high complexity) the better is the fit, and the smaller the hypothesis space (low complexity) the poorer is the fit. Hence, solving the model selection task is equivalent to considering a multi-objective optimization problem with two conflicting objectives.


\subsection{Multi-objective formulation and optimality}

A multi-objective optimization problem has two or more objectives which are conflicting. The objectives are supposed to be simultaneously optimized subject to a given set of constraints. These problems are commonly found in the fields of science, engineering, economics or any other field where optimal decisions are to be taken in the presence of trade-offs between two or more conflicting objectives. 

By interpreting the model selection problem as finding a trade-off between complexity and fit, we can formulate the following two objective problem where the objectives are jointly minimized. 
\begin{definition}[Multi-objective problem]\label{def:mop}
Let $\boldphi:\mathcal{H}\to\mathbb{N}\times\mathbb{R}$, $\boldphi=(\varphi_1,\varphi_2)$ denotes an objective vector, where 
\begin{itemize}
\item[(i)]the first objective $\varphi_1:\mathcal{H}\to\mathbb{N}$, $\varphi_1(f)=\min\{d\in \mathbb{N}:f\in\mathcal{H}_d\}$ represents the complexity of a model in terms of the number of variables; and
\item[(ii)]the second objective $\varphi_2:\mathcal{H}\to\mathbb{R}$ is the empirical risk $\varphi_2(f)=\frac{1}{n}\sum_{i=1}^nL(f(\mathbf{X}_i),Y_i)$, with quadratic loss function $L(f(\mathbf{X}_i),Y_i)=(Y_i-f(\mathbf{X}_i))^2$. This is same as the mean squared error. Some other suitable objective function may also be considered,  for instance, refer Section \ref{sec:genError}.
\end{itemize}
Then the optimization problem is given by
\begin{equation}
\begin{array}{ll}
\minimize_{f\in\mathcal{H}} & \boldphi(f) =
\left(\varphi_1(f),\varphi_2(f)\right), \\
\mbox{subject to} 
& f\in C,
\end{array}
\label{eq:bilevel_multi_obj}
\end{equation}
where $C\subset \mathcal{H}$ is a constraint set that has been included for the sake of generality and is not used hereafter.
\end{definition}

Usually, multi-objective problems do not have a single optimal solution which simultaneously maximizes or minimizes all of the objectives together; instead there is a set of solutions which are optimal in the sense that they are not dominated by any other solution. Once the models with best-fit corresponding to different complexities are available, the user could make the choice for the most preferred model.
\begin{definition}[Dominance]
A model $f^{(1)}$ is said to dominate the other model $f^{(2)}$, denoted as $f^{(1)}\succ f^{(2)}$, if both conditions 1 and 2 are true:
\begin{enumerate}
\item The model $f^{(1)}$ is no worse than $f^{(2)}$ in both objectives, or $\varphi_j(f^{(1)})\leq\varphi_j(f^{(2)})$, $j=1,2$.
\item The model $f^{(1)}$ is strictly better than $f^{(2)}$ in at least one objective, or $\varphi_j(f^{(1)})<\varphi_j(f^{(2)})$ for at least one $j\in\{1,2\}$.
\end{enumerate} 
\end{definition}



The concept of dominance gives a natural interpretation for optimality in multi-objective problems, because the quality of any two solutions can be compared on the basis of whether one point (model) dominates the other point (model) or not. 
\begin{definition}[Non-dominated set and Pareto-optimality]\label{def:optimality}
Among a set of solutions $\mathcal{P}\subset\mathcal{H}$, the non-dominated set of solutions $\mathcal{P}^{\star}$ are those that are not dominated by any member of the set $\mathcal{P}$, i.e. 
$$
\mathcal{P}^{\star}=\{f\in\mathcal{P}\ | \ \nexists g\in \mathcal{P} \ : \ g\succ f \}.
$$
When the set $P$ is the entire search space, i.e. $P=\mathcal{H}$, the resulting non-dominated collection of models $P^{\star}$ is called the Pareto-optimal set $\mathcal{H}^{\star}$.
\end{definition}

\section{Review on Model Selection Methods}\label{sec:overview}
A number of model selection criteria and methods have been suggested in the recent literature on statistical modeling and machine learning. However, given the lack of any clear standard, none of the methods has become dominant, and this leaves the user puzzled as to which approach to use. Most of the times, each of these selection criteria or methods lead to a different solution which makes it difficult for the user to pick a model. 
In this section, the existing models have been roughly categorized into different groups. It concludes with a summary of the central differences between these methods and the multi-objective framework suggested in this paper.

\subsection{Selection by Complexity Regularization}
Talking about the penalized model selection criteria, it can be found that there exist a number of model selection criteria in the literature. However, there are two criteria which are very commonly used. One is an information-theoretic method pioneered by \cite{ms-aic}, known as the Akaike Information Criteria (AIC) and the other one uses the Bayesian evidence, known as the Bayesian Information Criteria (BIC) (\cite{ms-bic}). A model which gives the least value for the criterion is the most preferred one. There are many other information criteria which are not commonly used and have been derived using similar principles as AIC and BIC. They are Deviance Information Criteria (DIC) (\cite{ms-dic}), Expected Akaike Information Criteria (EAIC), Fisher Information Criteria (FIC)~(\cite{ms-fic}), Generalized Information Criteria (GIC) (\cite{ms-gic}), Network Information Criteria (NIC) (\cite{ms-nic}), Takeuchi Information Criteria (TIC) (\cite{ms-tic}) and adaptive model selection (\cite{jasa-ams}). 

The classical information criteria can be essentially viewed as various forms of complexity regularization scheme, where the purpose is to penalize complex models based on their information content or using prior knowledge. In general, the choice of model by complexity regularization can be understood as solving a single objective minimization problem,
\begin{equation}\label{eq:penalty}
\hat{f}_n=\argmin_{f\in\mathcal{H}}\left\{J_{\lambda}(f):=\hat{R}_n(f)+ \lambda C(f)\right\}
\end{equation}
where $\hat{R}_n:\mathcal{H}\to\mathbb{R}$ denotes the empirical risk (e.g the function $\varphi_2$ in Problem~\ref{def:mop}), and $C:\mathcal{H}\to\mathbb{R}$ represents the cost of model which is commonly expressed in terms of the model size and sample size. 



\subsection{Stepwise Selection Methods}

Stepwise methods are commonly used to select the variables in a regression model. The methods commonly used are forward selection, backward elimination and stepwise regression. Forward selection method adds variables to the model until no remaining variable (outside the model) can add anything significant to the dependent variable. Forward selection starts with no variable in the model. Backward elimination is opposite to forward selection where variables are deleted one by one from the model until all remaining variables contribute something significant to the dependent variable. Backward elimination begins with a model which includes all the variables. Stepwise regression is a modification of the forward selection method in a way such that variables once included in the model are not guaranteed to stay. Stepwise method has a number of weaknesses, like, it usually results in models having too many variables, it suffers under collinearity, it is based on methods intended to test pre-specified hypothesis etc. A detailed discussion on these approaches and their weaknesses can be found in a recent study by \cite{ms-ratner}.

\subsection{Best Subset Method}
In the best subset method, usually an exhaustive or a branch and bound algorithm (\cite{bnb-algo}) is used to find the best models corresponding to fixed number of variables. The best subset selection finds the model with the greatest goodness-of-fit, for a fixed number of variables. When repeated for different number of variables, this procedure yields a set of Pareto-optimal models similar to what we are aiming for. The algorithm to find the best subset becomes computationally very expensive with increasing number of variables and is not a viable technique when the number of variables are very high. The MOGA-VS approach could be a useful strategy to produce the best-subset for large scale problems, when the conventional method fails or becomes very expensive. Many of the existing best subset implementations contain an upper bound on how many variables they can handle.

\subsection{Genetic algorithms and other heuristics}
A number of studies use genetic algorithms (GA) and other heuristic algorithms to choose regressors in a regression problem. Some of the studies to the knowledge of the authors are \cite{reg-heuristic1,reg-heuristic2, reg-heuristic3}. However, they differ from the method proposed in this paper as they assume a single objective function (usually an information criteria) and then use the heuristic algorithm to find an optimal regression model which optimizes the chosen objective. The Parallel Genetic Algorithm (PGA) framework suggested by \cite{zhu06} searches for an ensemble of good models, and then uses the entire set for subsequent model selection. A recent heuristic by \cite{wolters11} proposes a non-convergent approach for generating a large number of models for a fixed model size. Thereafter, a feature extraction problem is solved to choose the most appropriate model. The two studies are similar to ours, as they search for multiple models before accepting a particular model. However, they differ from MOGA-VS, as they do not target the entire range of Pareto-optimal models. In the process of converging towards the Pareto-optimal models, MOGA-VS also provides a high number of dominated models close to the Pareto frontier as a by-product of the optimization scheme.

\subsection{Bayesian Model Averaging}
An alternative to frequentist approaches for model selection is the use of techniques developed for Bayesian model averaging (BMA)~(\cite{bma-leamer,madigan94,bma-chatfield,hoeting99,montgomery10,clyde10}) technique for model selection. In our experiments, we consider one such method where BMA is used to rank models and uses the best subset method. The BMA technique computes the full joint posterior distribution over models which allows incorporation of model uncertainty in posterior inferences. 


A common strategy in BMA is to select the highest posterior probability model. As discussed by \cite{clyde10}, there are several other strategies to perform optimal model selection e.g. based on maximization of posterior expected utility. However, the difficulty in BMA is that when a large number of variables is involved, enumeration of the models in the hypothesis space $\mathcal{H}$ becomes a heavy task. Therefore, the use of Markov Chain Monte Carlo techniques or adaptive sampling is necessary even for problems of moderate size. Bayesian model averaging technique could also be used with our algorithm for selecting the best model from the non-dominated set of models.

\subsection{Central differences and motivation for MOGA-VS}

Both classical and multi-objective approaches have their pros and cons. The classical scheme is optimal if the chosen penalty scheme is a good representation of the user's preferences for trade-off between empirical risk and model complexity. However, many times, the model selection can turn out to be quite sensitive to the choice of complexity penalty. Furthermore, as discussed by~\cite{montgomery10}, uncertainty about the correct model specification can be very high in practical applications. For instance, in social sciences such as political research, where large sets of control variables are involved, an attempt to find a single best model is often poorly justified.

The multi-objective framework proposed in the present paper differs from the classical model selection techniques in the following respects:
\begin{itemize}
\item[(i)] {\it Multiple optimal solutions:} By treating the model selection task as a multi-objective optimization problem, we are always looking for a collection of Pareto-optimal solutions instead of attempting to choose one single optimal point directly. The Pareto set contains the best solution in terms of goodness-of-fit for each complexity. Therefore, these optimal solutions guarantee that for a given number of variables, there cannot exist a model which can provide a better fit for the training data.
\item[(ii)] {\it Separation of concerns:} The purpose in multi-objective approach is to avoid making an a priori choice of a complexity penalty. To accomplish this, a distinction is made between stages which can be objectively decided and those which are more dependent on the user's preferences and the particular application at hand. Finding the Pareto-optimal frontier is an optimization problem that can be solved without any a priori assumptions, whereas the choice of the preferred point(s) from the Pareto-optimal set is both preference as well as application dependent question. Therefore, in the proposed approach, the optimization stage, and decision-making stage are treated separately. By doing so, the multi-objective technique enhances understanding of the trade-off and what separates the alternative models.
\end{itemize}
The remaining question is how to find the Pareto-optimal solutions. Of course, for a finite search space $\mathcal{H}$, it is always possible to use brute force to find the Pareto-optimal set. However, such a naive approach would be intractable in practice. To solve the optimization problem in an efficient manner, our approach introduces a specialized multi-objective optimization framework that is based on evolutionary computation.

\section{The MOGA-VS Framework}\label{sec:moga}
In this section, we discuss a step-by-step procedure for the Multi-objective Genetic Algorithm for Variable Selection (MOGA-VS). The framework of this algorithm has been inspired by some of the existing evolutionary multi-objective (EMO) procedures (\cite{nsga2, spea2}). The presented algorithm has been specialized to handle the problem~\ref{def:mop} of variable selection efficiently. This section first provides a step-by-step procedure for the proposed algorithm (MOGA-VS). Then, the techniques used for visualizing the Pareto-optimal frontier and selection criteria are discussed.

\subsection{Step-by-Step Procedure for MOGA-VS}
Using the basic genetic algorithm framework, we suggest a specialized algorithm for producing the Pareto-optimal set of regression models when one objective is minimization of number of variables and the other objective is minimization of the in-sample mean squared error (other error measures may also be used, for example, minimization of generalization error is discussed in Section \ref{sec:genError}). It should be noted that whenever we refer to a population member it means we are referring to a regression model.  Each member (regression model) is represented by a binary string of the size of the total number of candidate variables. If a particular variable is present, the bit value is $1$; otherwise the bit value is $0$. For example, if there are $K$ candidate variables $(\boldx_1, \boldx_2, \ldots, \boldx_K)$ then the string $(1, 0, 0, 1, \ldots, 1)_K$ represents a regression model where the first variable is present, second is absent, third is absent, fourth is present and so on. Sum of the bits (number of variables present in the model) in the string represents the first objective and the mean squared error of the regression model represents  the second objective.

A step-by-step procedure for the Multi-objective Genetic Algorithm for Variable Selection (MOGA-VS) is described as follows:

\begin{enumerate}
\item[1.] Initialize a parent population, $\mathcal{P}$, of size $N$ by picking the regression variables randomly with probability 0.5 for each of the members.
\item[2.] Find the non-dominated set of solutions in the population, i.e. $\mathcal{P}^{\star}$.\footnote{Non-dominated members from a particular set could be identified by performing pairwise comparisons between all the members and selecting the ones which are not dominated by any member.}
\item[3.] Pick up any member from the non-dominated set $\mathcal{P}^{\star}$ and another member randomly from $\mathcal{P}$ to perform a single point crossover (\cite{goldberg-book}) of the binary strings leading to two offsprings. Repeat the process with different parents until $\lambda$ offspring members are produced. Add the offspring members to the set $\mathcal{O}$.
\item[4.] Perform a binary mutation (\cite{goldberg-book}) on each of the offspring members in set $\mathcal{O}$ by flipping the bits with a particular probability.
\item[6.] Add all the offspring members from the set $\mathcal{O}$ to $\mathcal{P}$. The size of $\mathcal{P}$ exceeds $N$, therefore delete dominated members with highest number of variables until the size of $\mathcal{P}$ becomes equal to $N$. In case all the members are non-dominated, then delete the members with highest number of variables.
\item[7.] If specified number of iterations, $i$, are done then terminate the process and report the non-dominated set from $\mathcal{P}$ else go to step 3.
\end{enumerate}

Choosing non-dominated parents for crossover, helps the algorithm in exploring members which are closer to the Pareto-optimal front. The output of the above algorithm is a non-dominated set of regression models $\mathcal{P}^{\star}$, which provides an approximation for the Pareto-optimal frontier $\mathcal{H}^{\star}$ of the entire hypothesis space. MOGA-VS does not require any additional parameters apart from the common genetic algorithm parameters like population size, number of iterations, number of offsprings, probability of crossover and probability of mutation. For these parameters one can follow the following guidelines,
Population size: $N=K$,
Crossover probability: $p_c=0.9$,
Mutation probability: $p_m=1/K$,
No. of offsprings: $\lambda=N$.
The algorithm should be executed for sufficient number of iterations, $i$, such that the non-dominated frontier achieved by the algorithm no longer improves. The population size should not be chosen less than $K$, otherwise the algorithm will not be able to approximate the entire Pareto-optimal frontier. For crossover and mutation probability, the commonly used parameters have been chosen. The performance of the algorithm is not susceptible to variation in these parameters. Once a set of trade-off models are obtained using the MOGA-VS procedure, the frontier needs to be examined to find the most preferred model. This can be done using a combination of graphics (Section~\ref{sec:moga-graphics}) and simple selection metrics (Section~\ref{sec:moga-select}).

\subsection{Visualizing the Pareto-optimal frontier}\label{sec:moga-graphics}

In order to get a quick overview of the obtained solutions, a commonly applied strategy is to construct an illustration of the Pareto-optimal set in the objective space. In a bi-objective framework two types of graphs can be considered:
\begin{itemize}
\item[(i)] {\it  Objective Space (OS)-plot:} The Pareto-optimal frontier obtained as a solution to Problem~\ref{def:mop} is a plot where the empirical risk $\varphi_1$ of the optimal models is presented as a decreasing function of model complexity, i.e. $\{(\varphi_1(f),\varphi_2(f))\in\mathbb{N}\times\mathbb{R} : f\in\mathcal{H}^{\star}\}$. The plot can be used for analyzing the trade-off between empirical risk and complexity before choosing one of the models. (see Section~\ref{sec:moga-select}). 
\item[(ii)] {\it Hypothesis Space (HS)-plot:} To get an idea on the structure of the Pareto-optimal models, i.e. what variables and how many are contained in them, a quick remedy is to consider a HS-plot which is reminiscent of a Gantt-chart in the hypothesis space. In HS-plot, the y-axis shows the variables contained in the Pareto-optimal models, and x-axis shows the optimal models as ordered according to their complexity, i.e. if the input-space $\mathcal{X}$ has $p$-variables, HS-plot corresponds to the set $\{(\varphi_1(f), \mathbf{x}_{k,f}):k\in\{1,\dots,p\}, f\in\mathcal{H}^{\star}\}$ where $\mathbf{x}_{k,f}\in\{0,1\}$ is an indicator for whether $f$ has the $k$-th variable or not. Gray-color in the chart indicates presence of a variable.
\end{itemize}

Illustrations of the graph-based tools and their use are discussed in the light of experiment studies in Section~\ref{sec:results}.

\subsection{Selecting preferred models}\label{sec:moga-select}

The graphical representations of the Pareto-optimal frontier can be used in conjunction with other criteria to decide which of the optimal models to choose for further examination. Some of these strategies are discussed below.
\begin{itemize}
\item[(i)] {\it Knee-point strategy:} Observing a knee-point (\cite{knee-bechikh,knee-das,knee-branke,knee-deb-gupta}) in the OS-plot can be considered as an indicator for an optimal degree of model complexity. A ``knee'' is interpreted as a saturation point in terms of goodness-of-fit vs complexity, where further increase in model complexity yields only minor improvement in fit. This strategy usually works quite well in many practical problems despite its simplicity.
\item[(ii)] {\it Bayesian statistics:} Another strategy is to consider the use of Bayesian Model Averaging approach along the Pareto-optimal frontier only. This would allow the user to select more than one optimal model to perform statistical inference. For example, if $\mathcal{B}^{\star}\subset\mathcal{H}^{\star}$ is a neighborhood of models surrounding the knee-point of the optimal frontier, the user might want to combine several models to perform posterior inferences on a given quantity of interest $\Delta$, i.e. $p(\Delta|Y)=\sum_{f\in \mathcal{B}^{\star}}p(\Delta|f,Y)p(f|Y)$. This is an appropriate strategy in particular when the user has prior information.
\item[(iii)] {\it Information criteria:} The models along the optimal frontier can also be analyzed using various information criteria discussed in Section~\ref{sec:overview}. Applying various information criteria, to these optimal models, allows the user to ascertain the extent of agreement among different information criteria.
\item[(iv)] {\it F-tests:} In case the user finds that several of the Pareto-optimal models are worthy candidates for further evaluation, then non-nested F-tests or encompassing F-tests between the competing specifications can be considered. More details on non-nested testing can be found e.g. in ~\cite{davidson04}.
\end{itemize}

\section{Results}\label{sec:results}
We provide the results on three different datasets in this section. The evaluation of MOGA-VS is first performed on two simulated datasets for which the true models are known, and then we study the performance on a real dataset. In the first example, we demonstrate how the commonly used information criteria act as value functions in a two-objective space. In the second example which involves a more difficult dataset, we compare the MOGA-VS approach against state-of-the-art model selection approaches.  Thereafter, the procedure is evaluated on a recently published communities and crimes dataset\footnote{http://archive.ics.uci.edu/ml/datasets/Communities+and+Crime} within the United States. The purpose is to find the attributes that best explain the total amount of violent crimes per 100K population. The section provides a comparison of the MOGA-VS framework with other state of the art techniques.

\subsection{Simulated Example 1}
This example is a function selection problem in additive regression. It evaluates the MOGA-VS algorithm on a simple problem for which the true model is known. To begin with, we provide the procedure, which we used to construct the dataset. Thereafter, we discuss the results obtained from MOGA-VS in the light of a ``knee-point-analysis'' of the Pareto-optimal frontier, where the mean squared error of the models is plotted against the number of coefficients. Next, the performance of the Pareto-models is analyzed in terms of information criteria to find out which models would be chosen, had we used a single objective procedure of minimizing AIC or BIC. 
In all simulations, we have used the following parameter values for MOGA-VS: Population Size: $N=26$, Number of iterations: $i=200$, Crossover probability: $p_c=0.9$, Mutation probability: $p_m=1/K$, No. of offsprings: $\lambda=N$.

We create a dataset which has five independent variables, $(x_1, x_2, x_3, x_4, x_5)$, and one dependent variable, $y$. The true regression model has
linear coefficients but the dependent variable $y$ can be a non linear function of the independent variables.
Now, we generate a row of independent variables and the dependent variable as described below:
\begin{equation}
\begin{array}{c}
x_1 \in rand(0,1),
x_2 \in rand(0,2), 
x_3 \in rand(0,1),
x_4 \in rand(0,4),
x_5 \in rand(0,5),\\
y = 10 + 5 x_1 + 2 e^{x_2} + 5 x_3 + 3 x_{3}^3 + 0.1 x_{4}^3 + 0.2 norm(0,1).\\
\end{array}
\label{eq:generate}
\end{equation}
here, $rand(a,b)$ represents a random number between $a$ and $b$ and $norm(0,1)$ is a normally distributed random number with zero mean and a standard deviation of one.
This operation gives us a single row of the dataset $\{y, x_1, x_2, x_3, x_4, x_5\}$. Repeating the operation $n$ number of times
we generate a dataset with $n$ number of rows. For this example, we have taken $n=1000$.
This dataset is given as input to the algorithm along with information about the possible functional forms. 
The possible functional forms for the 
independent variables $x_i, i \in \{1, 2, 3, 4, 5\}$ are $\{x_{i}, x_{i}^2, x_{i}^3, log (x_{i}), e^{x_{i}}\}, i \in \{1, 2, 3, 4, 5\}$. 
Using this information the algorithm creates a new dataset with $25$ columns. The functional forms of each of the variables are treated 
as separate variables.
The largest regression model will therefore have $26$ coefficients, one for each of the functional forms and one for the constant term in the model. When
the algorithm is executed we obtain regression models with minimum mean squared error for fixed number of regression coefficients. In other words, 
$26$ different models are produced as output with $k \in \{1, 2, 3, \ldots, 26\}$ number of regression coefficients in each of the models. 

\subsubsection{Knee Point Analysis}\label{sec:knee-point-experiment1}
\begin{figure*}[hbt]
\begin{minipage}[t]{0.45\linewidth}
\begin{center}
\epsfig{file=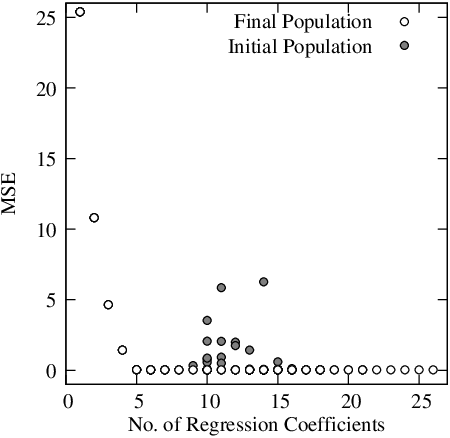,width=0.80\linewidth}
\end{center}
\vspace{-4mm}
\caption{Initial random models and the Pareto-optimal models.}
\label{fig:front-initial}
\end{minipage}\hfill
\begin{minipage}[t]{0.45\linewidth}
\begin{center}
\epsfig{file=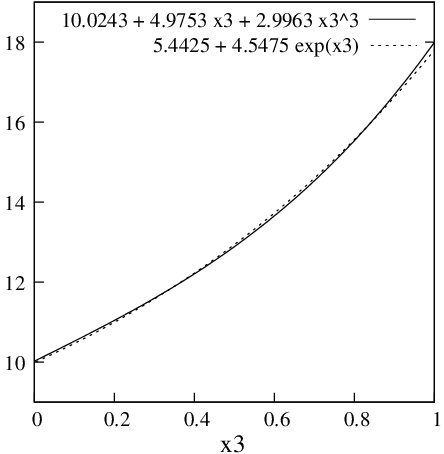,width=0.75\linewidth}
\end{center}
\vspace{-4mm}
\caption{Plot of the differing terms in the Pareto-optimal models with $6$ coefficients and $5$ coefficients.}
\label{fig:close}
\end{minipage}
\end{figure*}

MOGA-VS procedure is executed to generate a frontier of models shown in Figure~\ref{fig:front-initial}. The figure shows the initial random population (models) created by the algorithm and the final front which it achieved. It can be seen how over generations, the algorithm has progressed towards the front and has produced a diverse set of solutions. In this figure the model corresponding to 
``number of coefficients equal to $6$" represents exactly the $6$ variables present in the true model. The regression model for the dataset with true 
variables is as follows:
\begin{equation}
\begin{array}{c}
y = 10.0243 + 4.9926 x_1 + 2.0002 e^{x_2} + 4.9753 x_3 + 2.9963 x_{3}^3 + 0.0999 x_{4}^3\\
\end{array}
\label{eq:regress6}
\end{equation}

It is interesting to note, that the mean squared error falls sharply moving from model with $1$ coefficient to model with $5$ coefficients 
and does not change significantly thereafter. The model with $5$ coefficients visually appears a ``Knee Point" and represents a region of interest. The
model representing the true variables is expected to lie close to this region. A closer look reveals that the Pareto-optimal model with $5$ coefficients is:
\begin{equation}
\begin{array}{c}
y = 5.4425 + 4.9951 x_1 + 2.0001 e^{x_2} + 4.5475 e^{x_3} + 0.0999 x_{4}^3\\
\end{array}
\label{eq:regress5a}
\end{equation}

The difference in the Pareto-optimal models with $6$ and $5$ number of coefficients is primarily in the constant term and variable $x_3$.
The differing terms are $10.0243, 4.9753 x_3, 2.9963 x_{3}^3$ in the model with $6$ coefficients and 
$5.4425, 4.5475 e^{x_3}$ in the model with $5$ coefficients. Other terms are nearly equal in the two models. Figure~\ref{fig:close} shows the 
plot of summation of the differing terms for both the models in the 
range $[0, 1]$ ($x_3$ lies in this range). It can be seen that the two functions are significantly close in this range and therefore, the model with five coefficients also offers an acceptable fit.

\vspace{-3mm}

\begin{figure*}[hbt]
\begin{minipage}{0.48\linewidth}
\begin{center}
\epsfig{file=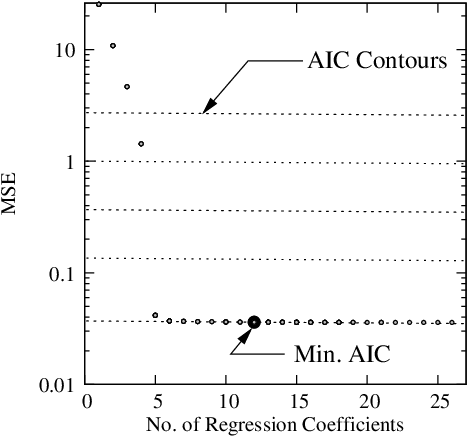,width=0.8\linewidth}
\end{center}
\caption{AIC value function (minimization) on the two objective plane.}
\label{fig:aic_vf}
\end{minipage}\hfill
\begin{minipage}{0.48\linewidth}
\begin{center}
\epsfig{file=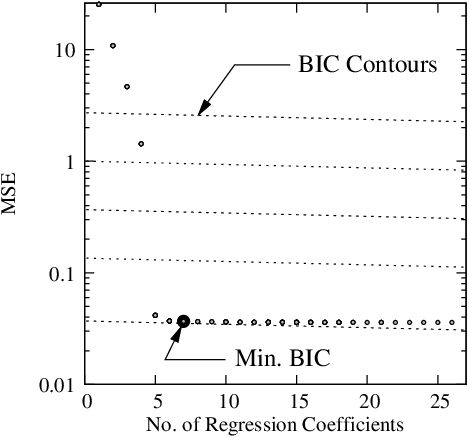,width=0.8\linewidth}
\end{center}
\caption{BIC value function (minimization) on the two objective plane.}
\label{fig:bic_vf}
\end{minipage}
\end{figure*}


\subsubsection{Analyzing Pareto-Optimal Regression Models Using Information Criteria}
Next, we compute the AIC and BIC values for all the frontier models obtained using MOGA-VS. 
We observe that the minimum AIC and BIC values correspond to models with $12$ (AIC: $-3.2962$) and $7$ (BIC: $-3.2550$) coefficients respectively. This leads us to the conclusion that an optimization done using AIC or BIC as an objective function will lead us to the models corresponding to $12$ or $7$ coefficients respectively. AIC and BIC act as a value function in a two objective case. We know that for normally distributed errors:
\begin{equation}
\begin{array}{c}
AIC=2k - 2\log(L), \\ 
BIC=k\log(n)-2\log(L),
\end{array}
\label{eq:aicbic}
\end{equation}
where $L$ is the likelihood of the parameters in the model.
In our case the first objective for minimization, $\varphi_1$, is number of variables, and
the second objective for minimization, $\varphi_2$, is mean squared error. This implies AIC and BIC can be written as:
\begin{equation}
\begin{array}{c}
AIC=2 \varphi_1 + n\log(\varphi_2), \\
BIC=\varphi_1 \log(n) + n \log(\varphi_2).
\end{array}
\label{eq:aicbic_2obj}
\end{equation}
In two objectives, these equations represent value functions which lead to a single
solution. Figure~\ref{fig:aic_vf} and \ref{fig:bic_vf} represent these value functions in the objective space, $\varphi_1$ and $\varphi_2$. Y-axis has been plotted on a log scale as the values for mean squared error are very close to each other after $\varphi_1=5$.

\subsection{Simulated Example 2}\label{sec:sim2}
We provide the results obtained from a simulated example with 100 variables and 500 observations. To increase the difficulty of the problem we have made all the 100 variables highly correlated by using the following mechanism:
\begin{equation}
x_i = 2z + \delta_i; \quad\quad i=1,2,\ldots,100, \quad\quad \delta_i,z \stackrel{iid}{\sim} N(0,1).
\end{equation}
This introduces a pairwise correlation among all the variables as $0.80$. The response variable is then constructed as follows:
\begin{equation}
y = 0.1x_1 + 0.2x_2 + 0.3x_3 + \ldots 1.0x_{10} + \epsilon; \quad\quad \epsilon \sim N(0,\sigma^2), \sigma=1.
\end{equation}
Once the response and predictor variables are generated, they are fed into the MOGA-VS algorithm with the following parameter values:
Population size: $N=100$,
Maximum number of iterations: $i=500$,
Crossover probability: $p_c=0.9$,
Mutation probability: $p_m=1/K$,
No. of offsprings: $\lambda=N$.

\begin{figure*}
\begin{minipage}[t]{0.49\linewidth}
\begin{center}
\epsfig{file=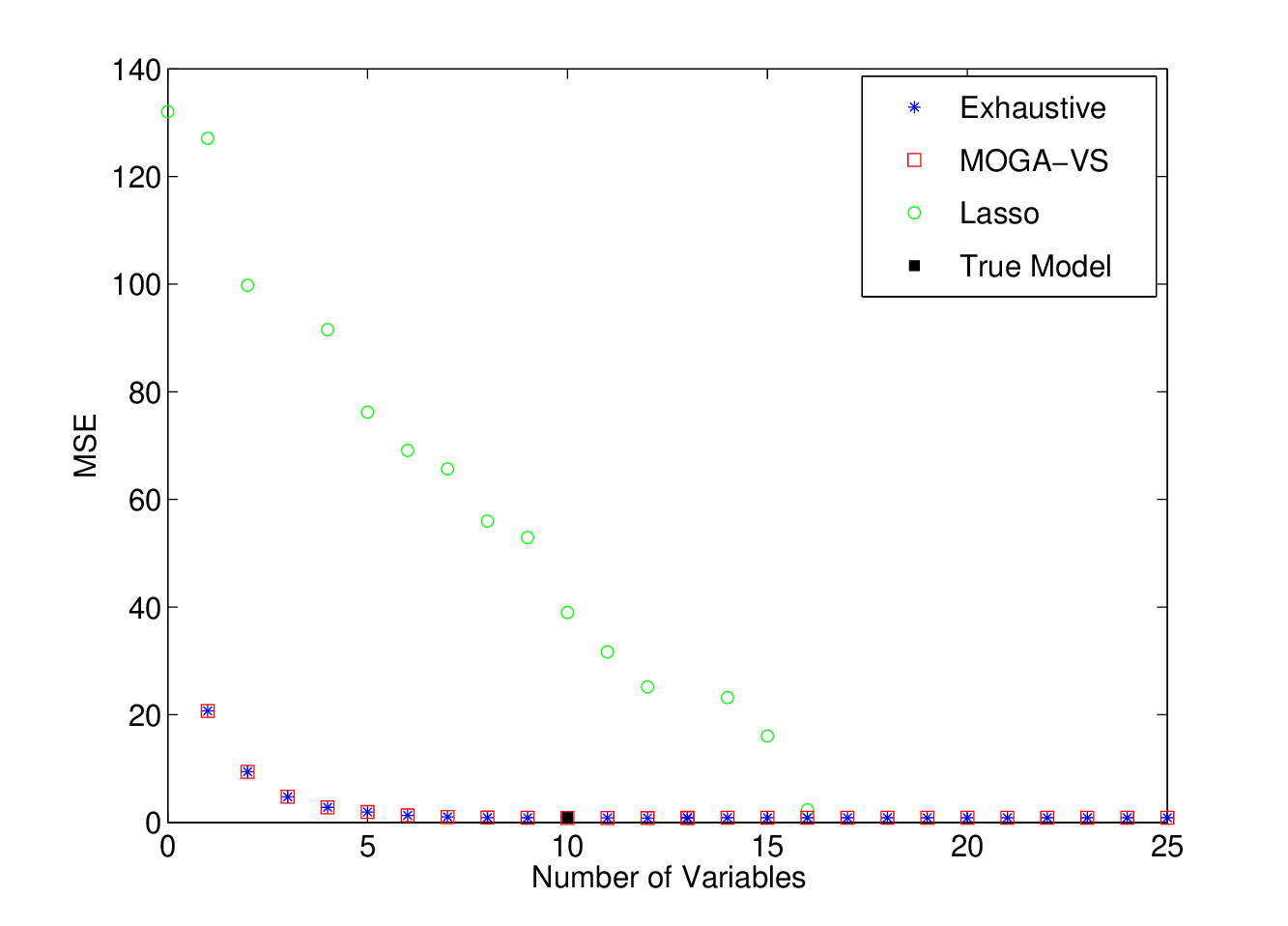,width=0.99\linewidth}
\end{center}
\caption{A part of the MOGA-VS frontier and a part of the Lasso frontier obtained using the simulated dataset from a sample run. The true model is also plotted.}
\label{fig:ex1-figure1}
\end{minipage}\hfill
\begin{minipage}[t]{0.49\linewidth}
\begin{center}
\epsfig{file=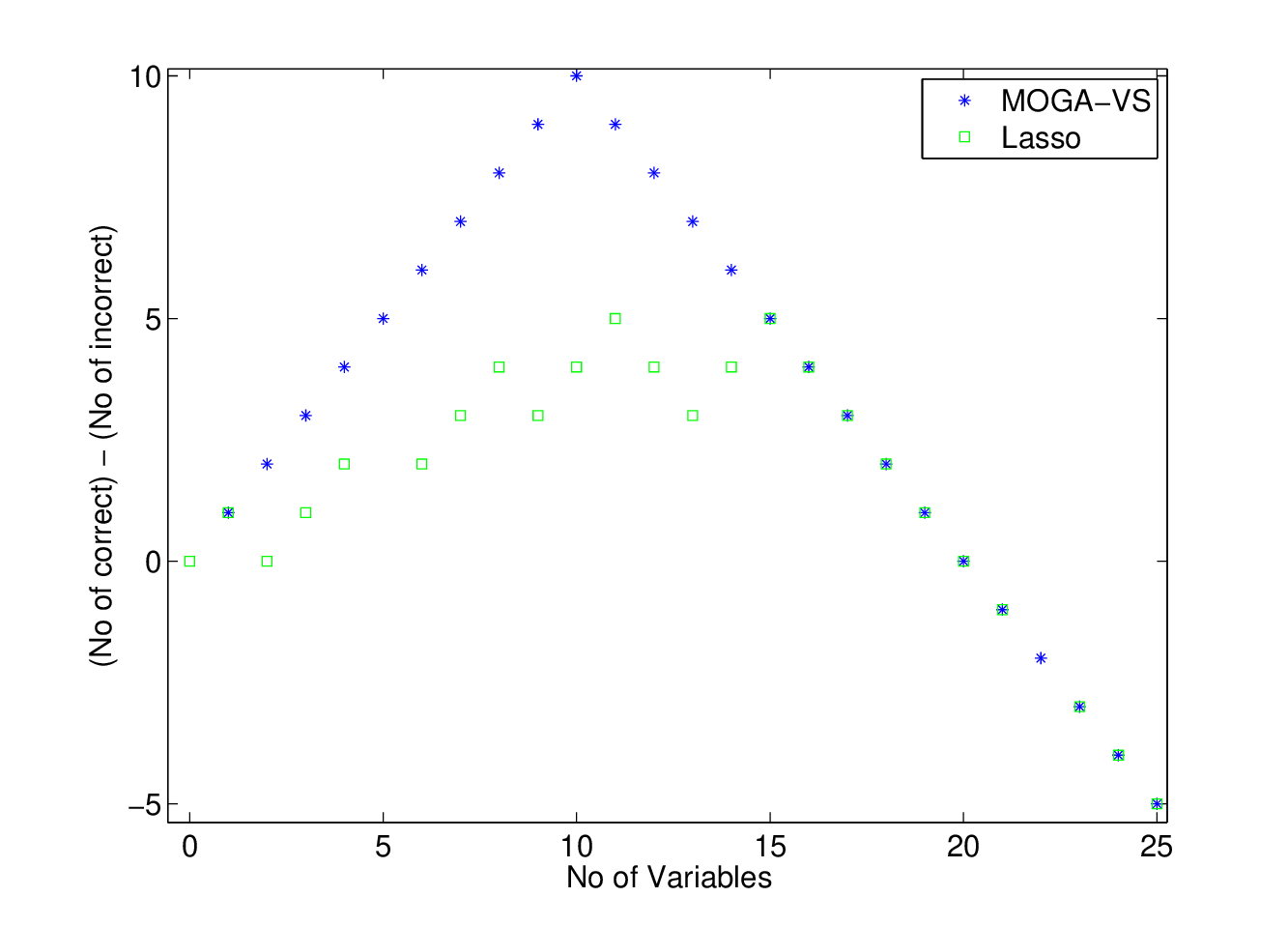,width=0.99\linewidth} 
\end{center}
\caption{The difference between the number of correct variables and the number of incorrect variables contained in the models produced by MOGA-VS and Lasso.}
\label{fig:ex1-figure3}
\end{minipage}
\end{figure*}

The algorithm produces a Pareto-frontier of models with complexities varying from 1 to 100. A part of the frontier produced by the algorithm is shown in Figure \ref{fig:ex1-figure1} for a dataset. We have performed a comparative study, where we examine the performance of our method against the Lasso (\cite{lasso}) scheme.
The Lasso frontier is generated by solving a number of single objective optimization problems with different parameter values\footnote{The Lasso parameter was incremented from 0 in steps of $0.01$, and a singe objective optimization problem was solved for each parameter until a model is obtained which includes all the variables.}. Figure \ref{fig:ex1-figure1} also shows the frontier obtained from the Lasso scheme using the same dataset. To evaluate the validity of the Pareto-optimal frontier obtained using MOGA-VS, we compare it with an exhaustive branch-and-bound search. An exhaustive branch-and-bound search was performed for complexities 1 to 25, and the results are shown in the same figure. One finds that the models obtained using MOGA-VS correspond to the models obtained using an exhaustive search.

Next, we evaluate the performance of the approaches in terms of number of correct or incorrect variables included in the various suggested models. We assign a value to each model as a difference of number of correct and incorrect variables included. For instance, the true model in this case will be assigned a value of 10, as it contains 10 correct and 0 incorrect variables. Any other model will have a value less than 10. Based on this, Figure \ref{fig:ex1-figure3} provides a plot of the difference values assigned to the models along the MOGA-VS and Lasso frontiers.

\begin{figure}
\begin{center}
\epsfig{file=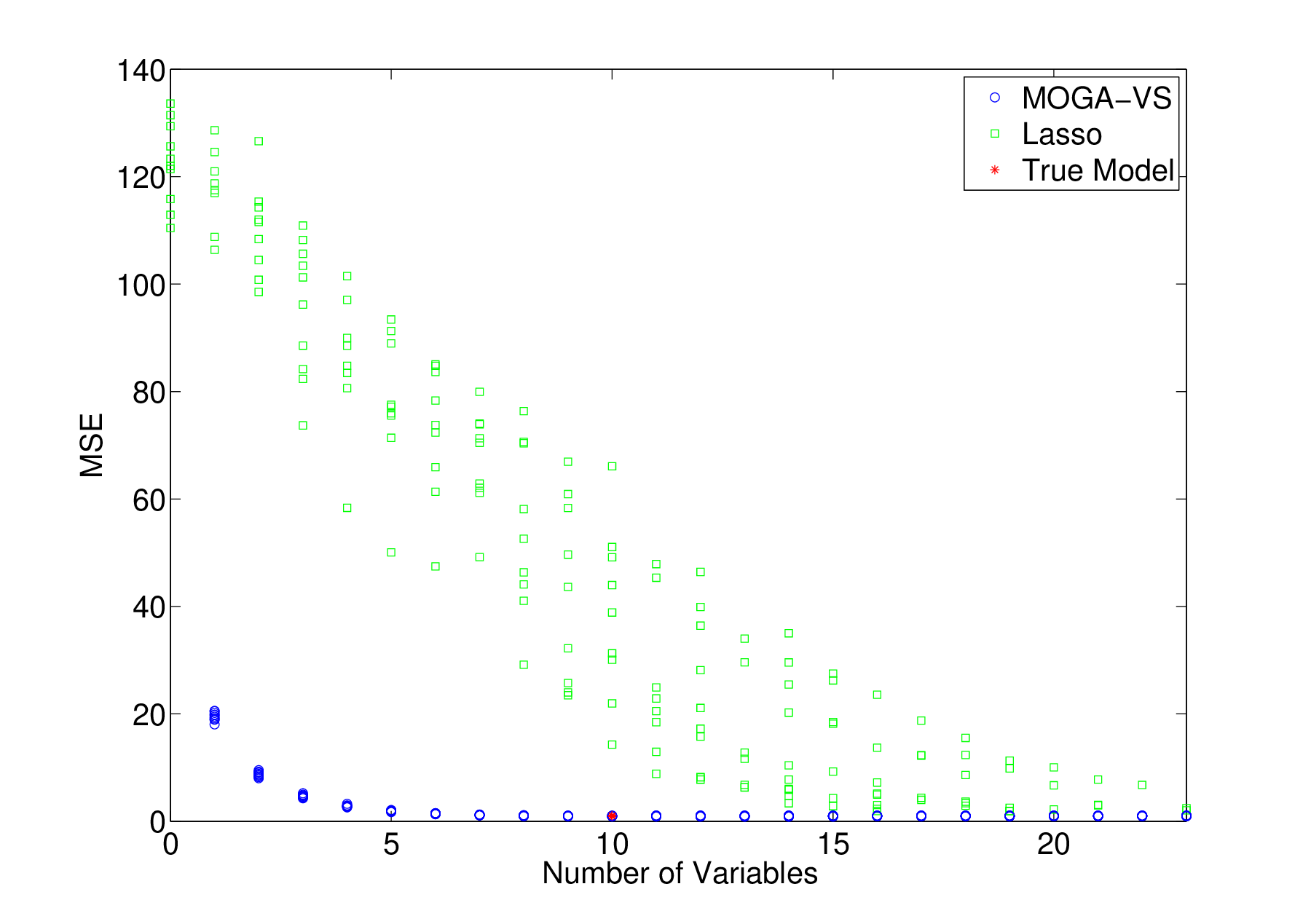,width=0.55\linewidth} 
\end{center}
\caption{The models obtained from 10 different runs of MOGA-VS and Lasso for the simulated dataset. The true model is also plotted.}
\label{fig:ex1-figure2}
\end{figure}

We have performed a simulation study where we execute each of the methods (MOGA-VS and Lasso) on 10 different datasets to observe the precision and accuracy. Figure \ref{fig:ex1-figure2} shows the results obtained from 10 sample runs of both the methods. It is easy to observe that the MOGA-VS scheme offers a high accuracy and precision, as the frontiers always pass close to the true model. Most of the models produced by Lasso are far away from the frontier. However, it should be noted that Lasso is not expected to produce models on the Pareto-optimal front. 

The results produced by MOGA-VS on this simulated example demonstrates its ability to explore the Pareto-optimal frontier consisting of trade-off models. It is noteworthy that in this example with highly correlated variables, the true model does not correspond to the knee region of the MOGA-VS frontier. This shows the inherent difficulty with the model selection task, which restricts one to rely entirely on one particular model selection scheme. Some of the existing methods like different information criteria, stepwise regression and Lasso have a strong foundation and are theoretically motivated. However, different models proposed by these methods highlight the subjective nature of the model selection task.


\subsection{Communities and crime}

The communities and crimes dataset (\cite{redmond09}) is formed as a combination of the socio-economic and law enforcement data from the 1990 US Census. The data also includes crime statistics from the 1995 FBI Uniform Crime Report. As discussed by \cite{redmond02}, the data set was originally collected to create a data-driven software tool called Crime Similarity System (CSS) for enabling cooperative information sharing among police departments. The idea in CSS is to utilize a variety of context variables ranging from socioeconomic, crime and enforcement profiles of cities to generate a list of communities that should be good candidates to co-operate due to their similar crime profiles. 

To demonstrate the performance of MOGA-VS framework, we consider the data-mining task of finding variables that best predict how many violent crimes are committed per 100K people. The number of candidate variables is 122, which corresponds to a hypothesis space $\mathcal{H}$ of size $2^{122}$. All of the variables have been normalized into $[0,1]$ interval to put all data into the same relative scale. The number of observations (or cases) is 1994, and each observation represents a single city or community. According to \cite{redmond02}, the variables have been chosen in close co-operation with police departments to find a collection of factors that provide a good coverage of the different aspects related to the community environment, law enforcement and crime. However, some of the variables included in the data set could not be used directly ``as is'' due to the large number of missing values. To alleviate this, imputation technique\footnote{The imputation was performed using the method {\em imputeData} (\cite{schafer97}) in the mclust-library on R.} was used to replace missing values on 20 attributes.

The MOGA-VS algorithm used the following parameter values:
Population size: $N=122$,
Maximum number of iterations: $i=500$,
Crossover probability: $p_c=0.9$,
Mutation probability: $p_m=1/K$,
No. of offsprings: $\lambda=N$.


\subsubsection{Analyzing the Pareto-optimal frontier}

\begin{figure}[t]
\begin{center}
\epsfig{file=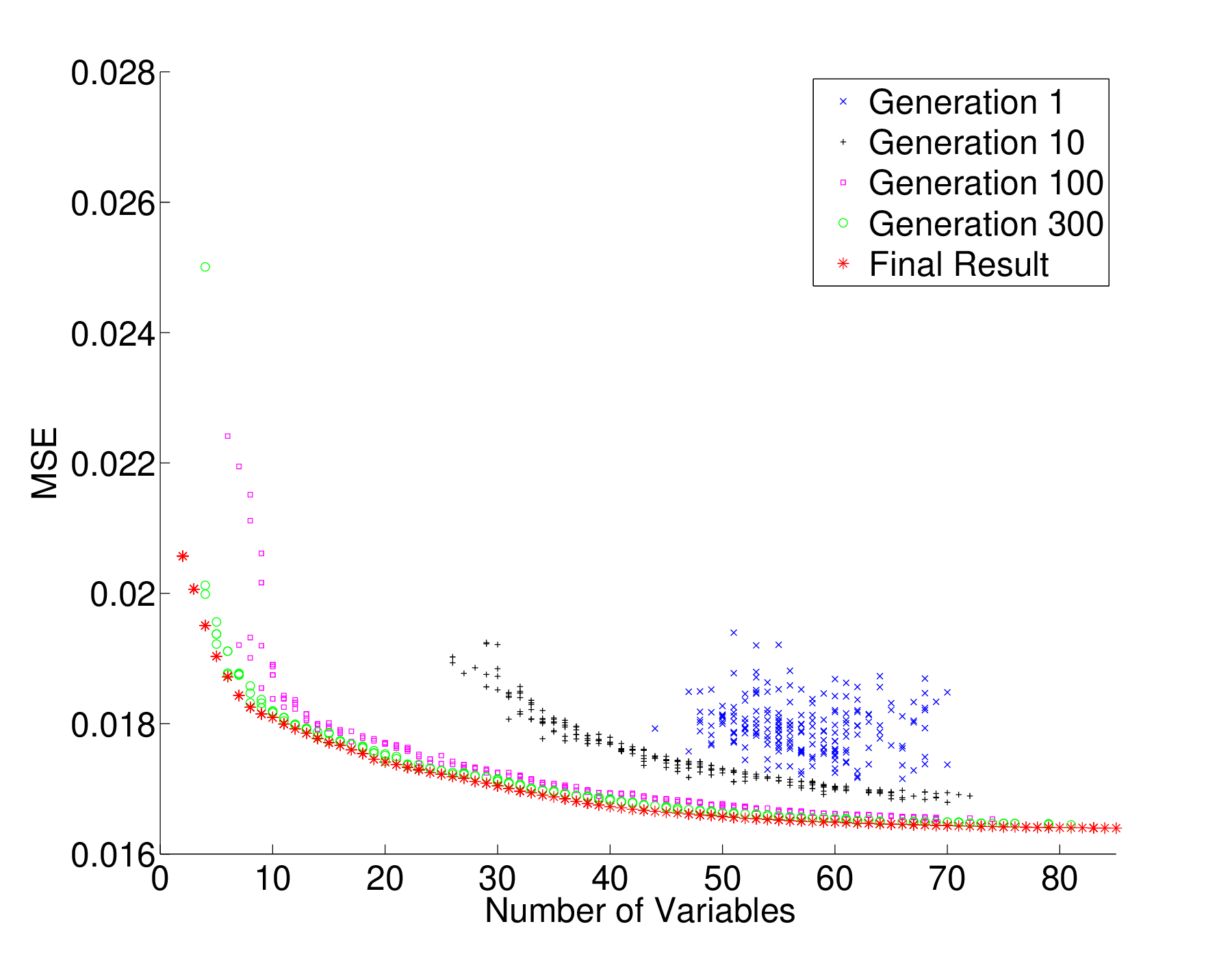,width=0.55\linewidth}
\end{center}
\caption{Pareto-optimal regression models: Mean squared error of the models has been shown on the y-axis and number of coefficients on the x-axis.}
\label{fig:front-crime}
\end{figure}

A description of the Pareto-optimal frontier is provided in Figure~\ref{fig:front-crime}. The plot shows the progress of the MOGA-VS algorithm when all the 1994 observations are considered. In addition to the final frontier, snapshots of intermediate generations are shown to illustrate the convergence towards the optimal front. The algorithm is able to provide a good approximation of the Pareto-frontier already by the 100th generation. However, more generations are needed to ensure convergence to the true frontier. The final result is the set of non-dominated solutions obtained by the algorithm after executing it for 500 generations. The plot for generation $1$ denotes the initial random models. It can be seen from the graph that the initial random models are initialized in the region close to 61 variables. The reason for this is that the initial bits are chosen to be either 0 or 1 with a $50\%$ probability. Therefore, the 122 bit chromosome has 61 expected variables. The algorithm is implemented on MATLAB, and required a total execution time of $37.27$ minutes on a Linux machine with 2.5 GHz Intel dual core processor and 2 GB of RAM. A total of 61,000 regression models were solved to arrive at the final frontier.

Most of the times we are interested in parsimonious models, so initializing the population with fewer 1s would boost the convergence. The convergence can be further enhanced by specifying constraints in the algorithm to perform a restricted search and producing only those models between $i$ to $j$ number of variables. Given a variable collection with 122 candidates, we would hardly want the final models to contain more than 20 variables. Therefore, a faster approximation of the interesting region of the Pareto-optimal frontier can be obtained by restricting the search for models with size between 1 to 20. Instead of starting the MOGA-VS algorithm with a random population, it is also possible to start the algorithm with close Pareto-solutions as initial population. One of the stepwise selection techniques could be first executed on the dataset to get the trajectory of the stepwise approach. Thereafter, the trajectory models could be used to generate the starting population for the MOGA-VS approach. Trajectory models could be a much better starting guess as compared with a random population. However, in this paper we do not use any starting guesses to justify that the MOGA-VS alone could lead to the set of Pareto-optimal solutions.

Based on visualization of the frontiers, we find that the knee of the curve lies in the region of 5 to 15 variables. The models which explain most of the variation in the response variable are the ones in the knee region. The incremental contributions of the remaining combinations of 112 variables are relatively small. This means that incorporating more explanatory variables would lead to only minor additional explanation of the variation. Choosing one of the models from the knee region offers a good compromise between goodness of fit and complexity. In the Table~\ref{tab:gant-chart} we provide the HS-plot which shows all non-dominated models with $5$ to $15$ number of variables produced by the MOGA-VS algorithm. The variables which are present in the model are marked as $1$ and the others are marked as $0$. This chart provides a useful information as to when the size of the model is increased by $1$ which variable(s) enter the model and which variable(s) are eliminated from the model. Consider a scenario, when a model size is increased from $k$ to $k+1$ causing one variable to leave the model and two variables to enter the model. It suggests that the explanatory power of the two variables entering the model is more than the explanatory power of the variable leaving the model when the remaining $k-1$ variables are kept intact. The chart helps a user to build an insight about the problem and enhances his understanding in order to choose a regression model wisely. After having the background information provided by the MOGA-VS algorithm, one can proceed to use a strategy for model selection. In the next sub-section we discuss the results obtained by other variable selection strategies.
\begin{table}[htp]
\begin{center}
\vspace{4mm}
\caption{Models in the knee region of the Pareto-optimal frontier}
\vspace{3mm}
\resizebox{.75\columnwidth}{!}{
\begin{tabular}{|l|c|c|c|c|c|c|c|c|c|c|c|}
\hline
racepctblack & \cellcolor[gray]{0.9}1 & \cellcolor[gray]{0.9}1 & \cellcolor[gray]{0.9}1 & \cellcolor[gray]{0.9}1 & \cellcolor[gray]{0.9}1 & \cellcolor[gray]{0.9}1 & \cellcolor[gray]{0.9}1 & \cellcolor[gray]{0.9}1 & \cellcolor[gray]{0.9}1 & \cellcolor[gray]{0.9}1 & \cellcolor[gray]{0.9}1 \\ \hline
PctIlleg & \cellcolor[gray]{0.9}1 & \cellcolor[gray]{0.9}1 & \cellcolor[gray]{0.9}1 & \cellcolor[gray]{0.9}1 & \cellcolor[gray]{0.9}1 & \cellcolor[gray]{0.9}1 & \cellcolor[gray]{0.9}1 & \cellcolor[gray]{0.9}1 & \cellcolor[gray]{0.9}1 & \cellcolor[gray]{0.9}1 & \cellcolor[gray]{0.9}1 \\ \hline
PctPersDenseHous & \cellcolor[gray]{0.9}1 & \cellcolor[gray]{0.9}1 & \cellcolor[gray]{0.9}1 & \cellcolor[gray]{0.9}1 & \cellcolor[gray]{0.9}1 & \cellcolor[gray]{0.9}1 & \cellcolor[gray]{0.9}1 & \cellcolor[gray]{0.9}1 & \cellcolor[gray]{0.9}1 & \cellcolor[gray]{0.9}1 & \cellcolor[gray]{0.9}1 \\ \hline
HousVacant & \cellcolor[gray]{0.9}1 & \cellcolor[gray]{0.9}1 & \cellcolor[gray]{0.9}1 & \cellcolor[gray]{0.9}1 & \cellcolor[gray]{0.9}1 & \cellcolor[gray]{0.9}1 & \cellcolor[gray]{0.9}1 & \cellcolor[gray]{0.9}1 & \cellcolor[gray]{0.9}1 & \cellcolor[gray]{0.9}1 & \cellcolor[gray]{0.9}1 \\ \hline
MalePctDivorce & \cellcolor[gray]{0.9}1 & \cellcolor[gray]{0.9}1 & \cellcolor[gray]{0.9}1 & \cellcolor[gray]{0.9}1 & \cellcolor[gray]{0.9}1 & \cellcolor[gray]{0.9}1 & \cellcolor[gray]{0.9}1 & \cellcolor[gray]{0.9}1 & \cellcolor[gray]{0.9}1 & \cellcolor[gray]{0.9}1 & \cellcolor[gray]{0.9}1 \\ \hline
pctWWage & 0 & \cellcolor[gray]{0.9}1 & \cellcolor[gray]{0.9}1 & \cellcolor[gray]{0.9}1 & \cellcolor[gray]{0.9}1 & \cellcolor[gray]{0.9}1 & \cellcolor[gray]{0.9}1 & \cellcolor[gray]{0.9}1 & \cellcolor[gray]{0.9}1 & 0 & 0 \\ \hline
pctUrban & 0 & 0 & \cellcolor[gray]{0.9}1 & \cellcolor[gray]{0.9}1 & \cellcolor[gray]{0.9}1 & \cellcolor[gray]{0.9}1 & \cellcolor[gray]{0.9}1 & \cellcolor[gray]{0.9}1 & \cellcolor[gray]{0.9}1 & \cellcolor[gray]{0.9}1 & \cellcolor[gray]{0.9}1 \\ \hline
NumStreet & 0 & 0 & 0 & \cellcolor[gray]{0.9}1 & \cellcolor[gray]{0.9}1 & \cellcolor[gray]{0.9}1 & \cellcolor[gray]{0.9}1 & \cellcolor[gray]{0.9}1 & \cellcolor[gray]{0.9}1 & \cellcolor[gray]{0.9}1 & \cellcolor[gray]{0.9}1 \\ \hline
numbUrban & 0 & 0 & 0 & 0 & \cellcolor[gray]{0.9}1 & 0 & 0 & \cellcolor[gray]{0.9}1 & \cellcolor[gray]{0.9}1 & \cellcolor[gray]{0.9}1 & \cellcolor[gray]{0.9}1 \\ \hline
RentLowQ & 0 & 0 & 0 & 0 & 0 & \cellcolor[gray]{0.9}1 & \cellcolor[gray]{0.9}1 & \cellcolor[gray]{0.9}1 & \cellcolor[gray]{0.9}1 & \cellcolor[gray]{0.9}1 & \cellcolor[gray]{0.9}1 \\ \hline
MedRent & 0 & 0 & 0 & 0 & 0 & \cellcolor[gray]{0.9}1 & \cellcolor[gray]{0.9}1 & \cellcolor[gray]{0.9}1 & \cellcolor[gray]{0.9}1 & \cellcolor[gray]{0.9}1 & \cellcolor[gray]{0.9}1 \\ \hline
MedOwnCost...Mtg & 0 & 0 & 0 & 0 & 0 & 0 & \cellcolor[gray]{0.9}1 & \cellcolor[gray]{0.9}1 & \cellcolor[gray]{0.9}1 & \cellcolor[gray]{0.9}1 & \cellcolor[gray]{0.9}1 \\ \hline
PctWorkMom & 0 & 0 & 0 & 0 & 0 & 0 & 0 & 0 & \cellcolor[gray]{0.9}1 & \cellcolor[gray]{0.9}1 & \cellcolor[gray]{0.9}1 \\ \hline
pctWSocSec & 0 & 0 & 0 & 0 & 0 & 0 & 0 & 0 & 0 & \cellcolor[gray]{0.9}1 & \cellcolor[gray]{0.9}1 \\ \hline
PctKids2Par & 0 & 0 & 0 & 0 & 0 & 0 & 0 & 0 & 0 & \cellcolor[gray]{0.9}1 & \cellcolor[gray]{0.9}1 \\ \hline
LemasSwFTFieldOps & 0 & 0 & 0 & 0 & 0 & 0 & 0 & 0 & 0 & 0 & \cellcolor[gray]{0.9}1 \\ \hline
\multicolumn{ 12}{|c|}{} \\ \hline
No. of Variables & 5 & 6 & 7 & 8 & 9 & 10 & 11 & 12 & 13 & 14 & 15 \\ \hline
MSE $\times$ 100 & 2.00 & 1.92 & 1.89 & 1.88 & 1.86 & 1.85 & 1.84 & 1.83 & 1.82 & 1.81 & 1.80 \\ \hline
\end{tabular}
}
\label{tab:gant-chart}
\end{center}
\vspace{-5mm}
\end{table}

\subsubsection{Results obtained from other techniques}
\begin{figure*}
\begin{minipage}[t]{0.49\linewidth}
\begin{center}
\epsfig{file=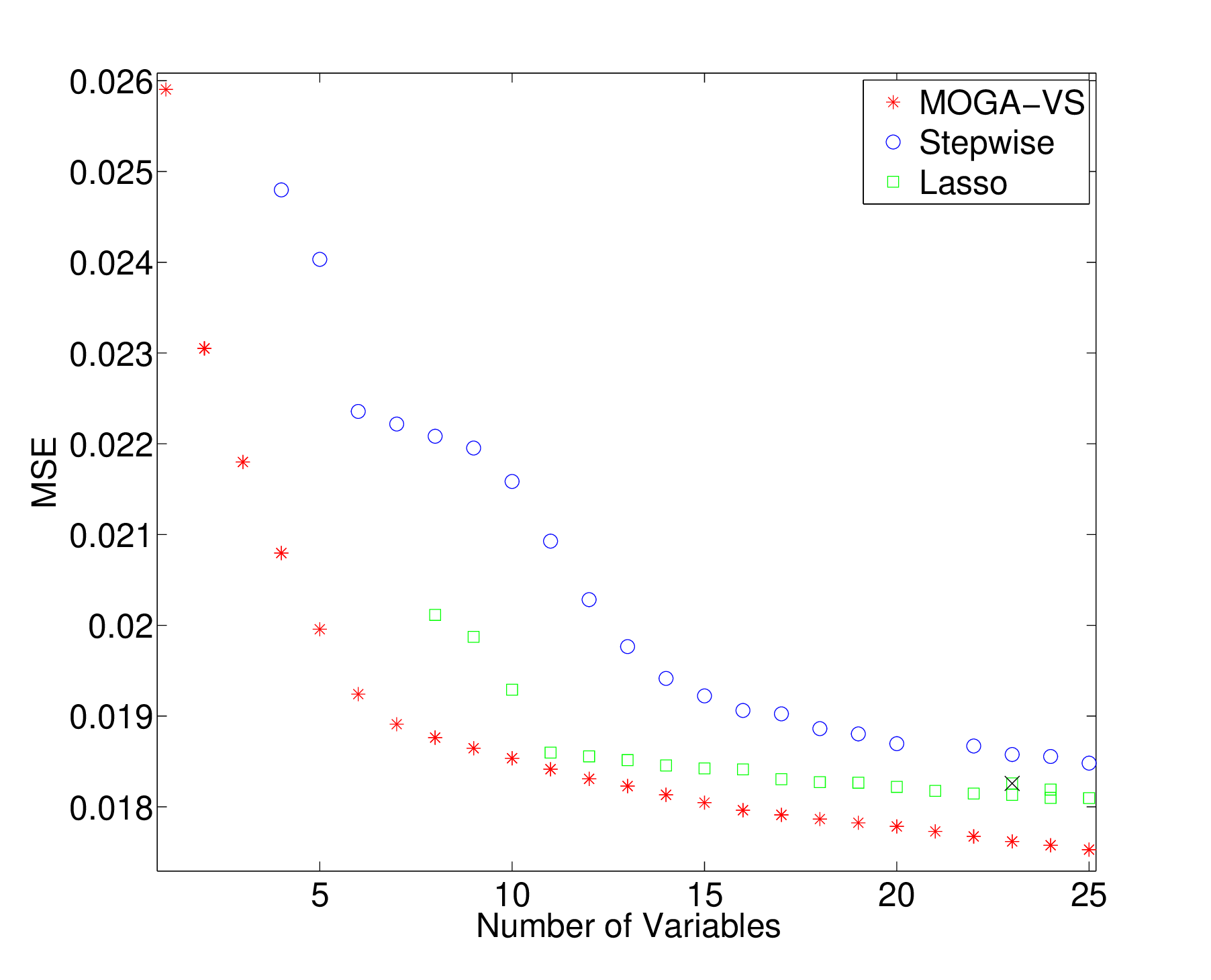,width=0.99\linewidth}
\end{center}
\caption{A part of the MOGA-VS frontier, a part of the Lasso frontier and Stepwise trajectory obtained using the entire communities and crime data as training set.}
\label{fig:figure2}
\end{minipage}\hfill
\begin{minipage}[t]{0.49\linewidth}
\begin{center}
\epsfig{file=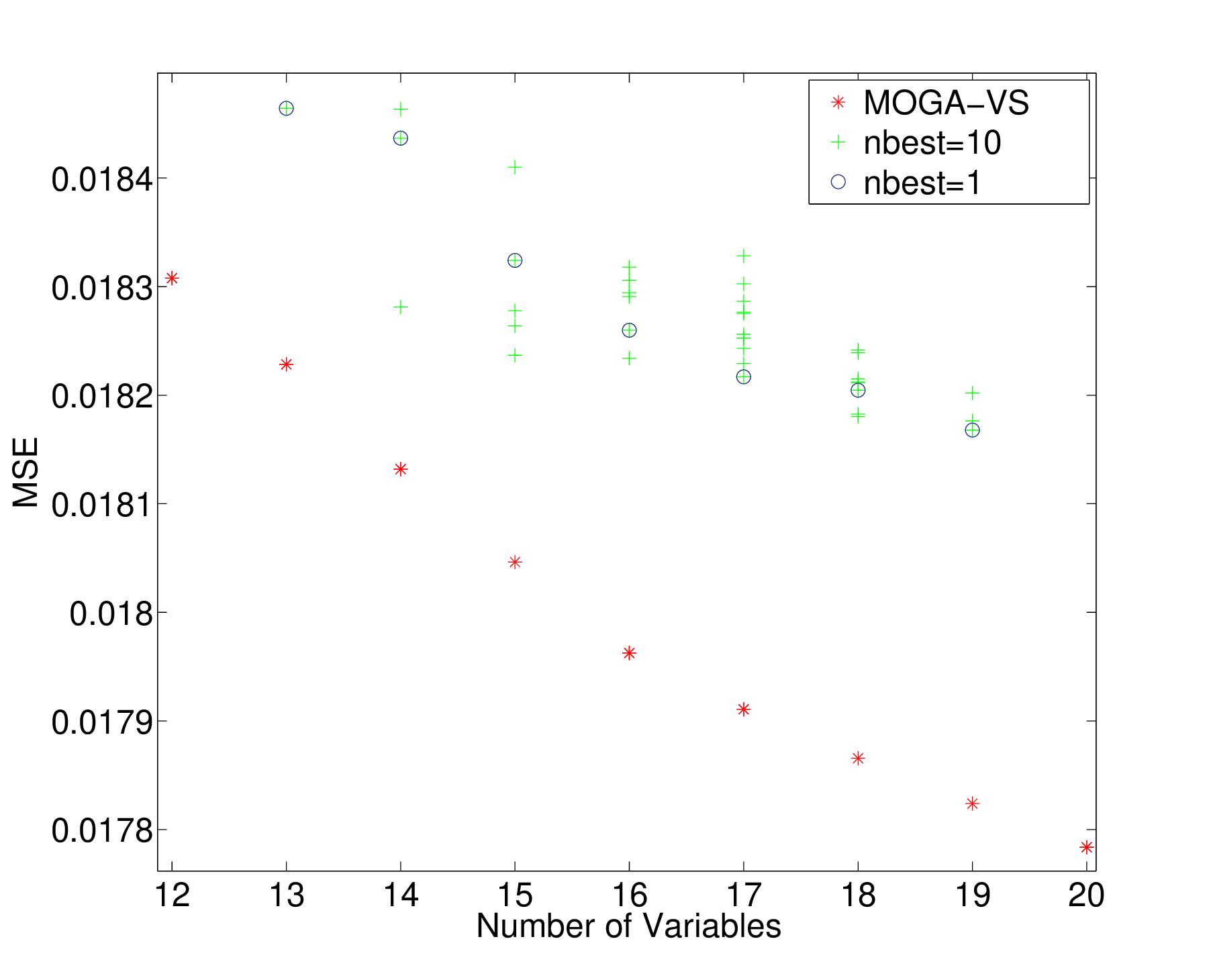,width=0.99\linewidth} 
\end{center}
\caption{A part of the MOGA-VS frontier and the Leaps results for two different parameters values obtained using the entire communities and crime data as training set.}
\label{fig:figure3}
\end{minipage}
\end{figure*}

In this section, we present the results from other state-of-the-art techniques used for variable selection. Figure~\ref{fig:figure2} shows the frontier obtained using MOGA-VS against the frontier produced by the  Lasso scheme of \cite{lasso}. The Lasso frontier is obtained by solving single objective optimization problems with different parameter values\footnote{The Lasso parameter was incremented from 0 in steps of $0.01$, and a singe objective optimization problem was solved for each parameter until a model is obtained which includes all the variables.}. Along with the two frontiers, the figure also shows the trajectory for a stepwise regression scheme, which is found to be close to the frontiers. The model shown with a cross is the final model chosen by the stepwise regression method. The initial points for Lasso and Stepwise method are not visible in the figure as they have a high MSE value. 
Figure~\ref{fig:figure3} shows the models obtained using two different parameter values for the Leaps algorithm\footnote{http://cran.r-project.org/web/packages/BMA/BMA.pdf}, i.e. $nbest=1$ and $nbest=10$. The parameter $nbest$ represents the number of models for each variable size to be generated by the leaps algorithm. The results produced in the first two figures are obtained by utilizing the entire data-set for training. In both figures, we observe that the models produced by Lasso, Stepwise and Leaps are dominated by the MOGA-VS frontier.

To examine the sensitivity of the model selection techniques for the choice of training and evaluation data, we proceed with another experiment, where the original data-set is divided into training and evaluation set. To obtain the average results, we create multiple test-sets of training and evaluation data by randomly choosing $50\%$ of the rows from the original data-set as training set and the remaining rows as evaluation set. Aggregated results of the randomization experiment are furnished in Tables~\ref{tab:simulation1} and ~\ref{tab:simulation2}, which provide a performance metric for all the methods across 20 different test sets of training and evaluation data.  For the $i^{th}$ test-set we generate the frontiers using one of the methods, and calculate the average MSE (say $\kappa_{i}^{method}$) for a part of the frontier models\footnote{The reason for considering only a part of the frontier is because not all the methods produce models across the entire frontier. Stepwise trajectory contains models from 1 to 25 variables and Leaps contains models from 6 to 20 variables.}. The comparison metric is computed by taking an average of $\kappa_{i}^{method}$ across 20 test sets (say $\kappa^{method}$) for each of the methods. Lower value for $\kappa^{method}$ denotes a better performance. We conclude from the results that the best-fit models for the training set perform better even on evaluation set, but this may not always be true. The performance metric denotes a slightly better performance for the MOGA-VS algorithm on the training sets as well as the evaluation sets.

\begin{table}[htp]
\caption{Values for $\kappa^{MOGAVS}$, $\kappa^{Lasso}$ and $\kappa^{Stepwise}$ computed from 20 frontiers. Models containing 1 to 25 number of variables were considered while taking the average.}
\label{tab:simulation1}
\begin{center}
\resizebox{.5\columnwidth}{!}{
    \begin{tabular}{|c|c|c|c|}
        \hline
        & MOGA-VS & Lasso  & Stepwise \\ \hline
	Training Set & 0.0184 & 0.0244 & 0.0218 \\
        Evaluation Set & 0.0204  & 0.0277 & 0.0233   \\
        \hline
    \end{tabular}
}
\end{center}
\end{table}

\begin{table}[htp]
\caption{Values for $\kappa^{MOGAVS}$, $\kappa^{Leaps(nbest=1)}$ and $\kappa^{Leaps(nbest=10)}$ computed from 20 frontiers. Models containing 6 to 20 number of variables were considered while taking the average.}
\label{tab:simulation2}
\begin{center}
\resizebox{.7\columnwidth}{!}{
    \begin{tabular}{|c|c|c|c|}
        \hline
        & MOGA-VS & Leaps (nbest=1) & Leaps (nbest=10) \\ \hline
	Training Set & 0.0181 & 0.0208 & 0.0195 \\
        Evaluation Set & 0.0195  & 0.0217        & 0.0206         \\
        \hline
    \end{tabular}
}
\end{center}
\end{table}

\begin{figure*}
\begin{minipage}[t]{0.49\linewidth}
\begin{center}
\epsfig{file=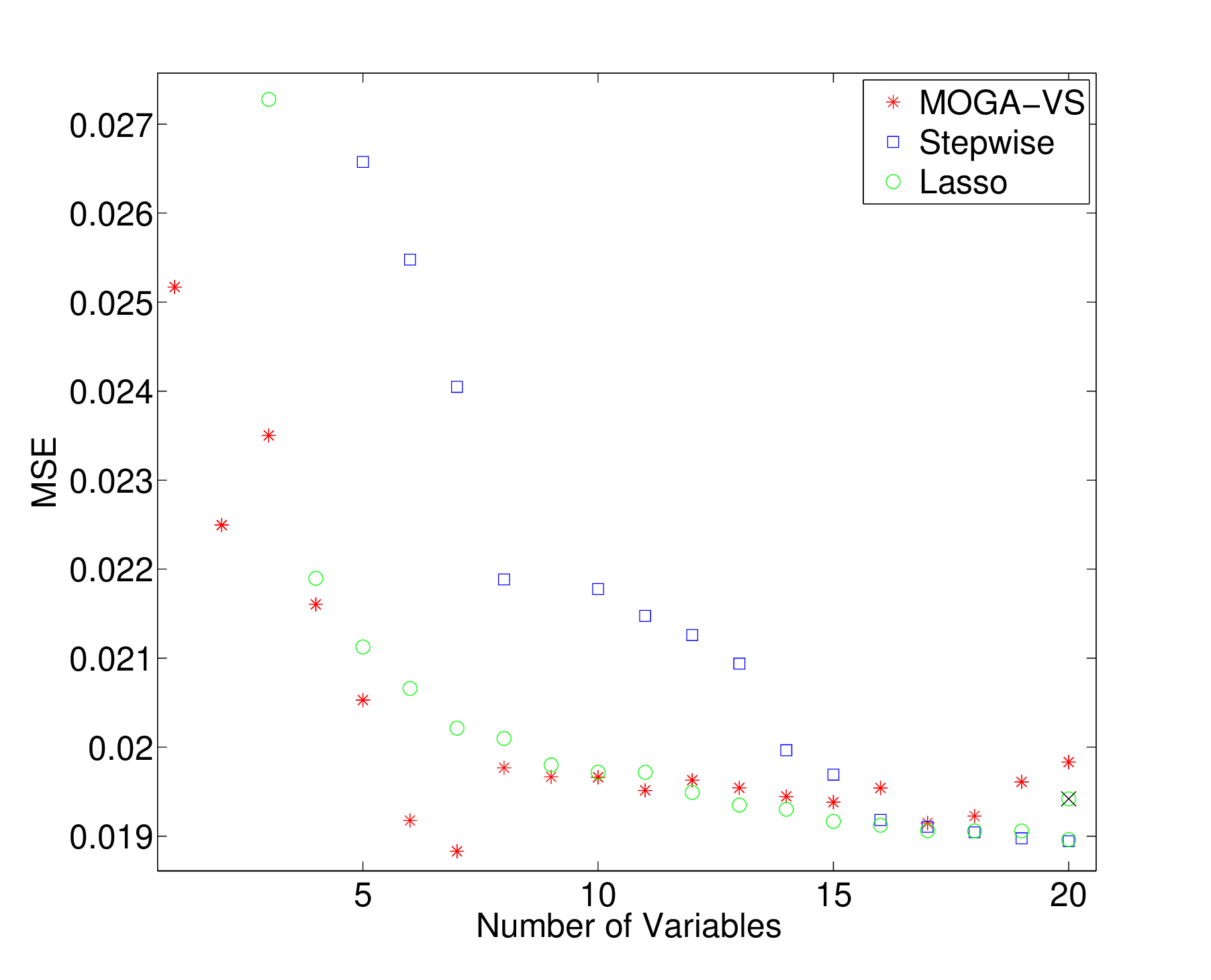,width=0.99\linewidth}
\end{center}
\caption{A part of the MOGA-VS frontier, Lasso frontier, and Stepwise trajectory on the evaluation set when $50\%$ of the communities and crime data is used as training set and the remaining $50\%$ as evaluation set.}
\label{fig:figure4}
\end{minipage}\hfill
\begin{minipage}[t]{0.49\linewidth}
\begin{center}
\epsfig{file=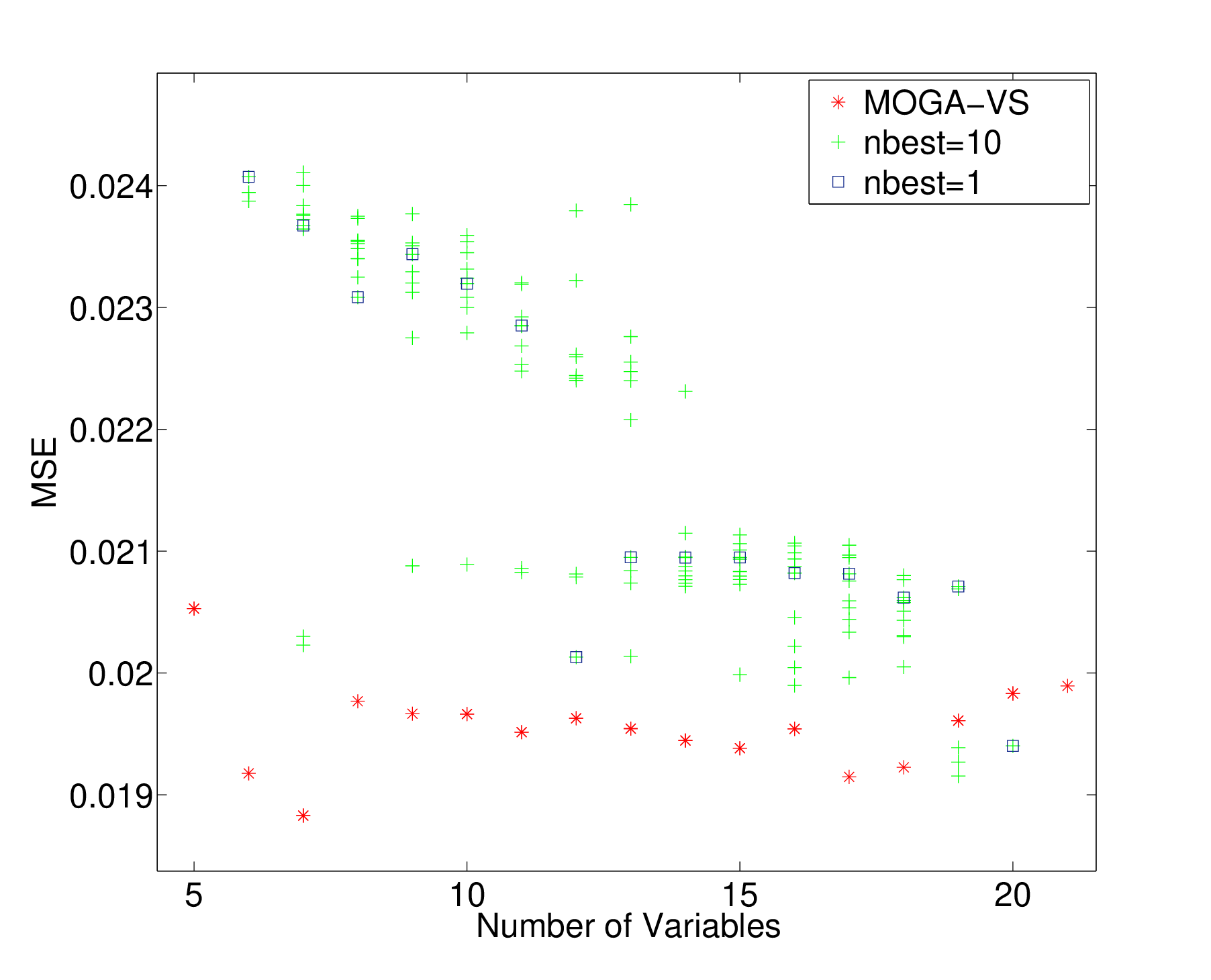,width=0.99\linewidth} 
\end{center}
\caption{A part of the MOGA-VS frontier, and Leaps results for two different parameters on the evaluation set when $50\%$ of the communities and crime data is used as training set and the remaining $50\%$ as evaluation set.}
\label{fig:figure5}
\end{minipage}
\end{figure*}

Figures~\ref{fig:figure4} and~\ref{fig:figure5} provide the results on the evaluation data-set for MOGA-VS, Stepwise Regression, Lasso and Leaps for a particular test-set out of 20 randomly generated test-sets. The cross-mark on Figure~\ref{fig:figure4} is the model suggested by the stepwise regression method. We can observe from the graphs that the frontier for MOGA-VS is slightly ahead of the other frontiers particularly towards models with smaller number of variables.
\begin{table}[htp]
\vspace{5mm}
\caption{Experiment Results: Model variables, + denotes a positive coefficient, - denotes a negative coefficient and no sign denotes that the variable is absent in the model}
\label{tab:experiment2-pga}
\begin{center}
\resizebox{.75\columnwidth}{!}{
\begin{tabular}{|l|c|c|c|c|c|c|c|} \hline
Variable	&	BMA-1	&	BMA-2	&	BMA-3	&	BMA-4	&	BMA-5	&	Stepwise	&	PGA	\\
\hline
racepctblack	&	$+$	&	$+$	&	$+$	&	$+$	&	$+$	&	$+$	&	$+$	\\
PctIlleg	&	$+$	&	$+$	&	$+$	&	$+$	&	$+$	&	$+$	&	$+$	\\
PctPersDenseHous	&	$+$	&	$+$	&	$+$	&	$+$	&	$+$	&	$+$	&	$+$	\\
HousVacant	&	$+$	&	$+$	&	$+$	&	$+$	&	$+$	&	$+$	&	$+$	\\
pctWWage	&		&	$-$	&	$-$	&		&		&		&		\\
pctUrban	&	$+$	&	$+$	&	$+$	&	$+$	&	$+$	&	$+$	&	$+$	\\
NumStreet	&	$+$	&	$+$	&	$+$	&	$+$	&	$+$	&	$+$	&	$+$	\\
numbUrban	&	$-$	&	$-$	&	$-$	&	$-$	&	$-$	&		&		\\
RentLowQ	&		&	$-$	&	$-$	&	$-$	&		&		&	$-$	\\
MedRent	&		&	$+$	&	$+$	&	$+$	&		&		&		\\
MedOwnCostPctIncNoMtg	&	$-$	&	$-$	&	$-$	&	$-$	&	$-$	&	$-$	&	$-$	\\
PctWorkMom	&	$-$	&	$-$	&	$-$	&	$-$	&	$-$	&	$-$	&	$-$	\\
PctKids2Par	&	$-$	&	$-$	&	$-$	&	$-$	&	$-$	&	$-$	&	$-$	\\
agePct12t29	&	$-$	&		&	$-$	&	$-$	&	$-$	&	$-$	&	$-$	\\
pctWInvInc	&	$-$	&	$-$	&	$-$	&	$-$	&	$-$	&	$-$	&	$-$	\\
PctEmploy	&	$+$	&	$+$	&	$+$	&	$+$	&		&		&		\\
MalePctNevMarr	&	$+$	&		&	$+$	&	$+$	&	$+$	&	$+$	&		\\
	&		&		&		&		&		&	$+11$ other	&		\\
No. of Variables	&	14	&	15	&	17	&	16	&	13	&	23	&	12	\\
MSE$\times100$ & 2.03 & 1.98 & 1.97 & 2.02 & 2.03 & 1.83 & 2.08 \\
Post. Prob.	&	0.295	&	0.2	&	0.168	&	0.141	&	0.074	&		&		\\
\hline
\end{tabular}
}
\end{center}
\end{table}

In Table~\ref{tab:experiment2-pga} we provide the results for BMA, PGA and stepwise regression methods. A direct comparison with the MOGA-VS results can be obtained by comparing Table~\ref{tab:experiment2-pga} against Table~\ref{tab:gant-chart}. The table shows the variables which are present in different models along with coefficient sign patterns. A plus sign indicates a positive coefficient for the variable, a negative sign indicates a negative coefficient for the variable, and no sign indicates that the variable is absent in the model. For BMA we have presented the top $5$ models ranked by posterior probabilities. The best model proposed by BMA and the model proposed by PGA has 14 and 12 number of variables respectively. Both of these models lie close to the knee region of the Pareto-frontier. On the other hand Stepwise regression proposes a model with $23$ variables which could be rejected. A closer look at the table shows that the models proposed by BMA and PGA agree with each other and contain mostly common variables. If the user wants lesser complex model with less than 12 variables, then the models in the knee region of the Pareto-frontier offer relevant alternatives. Finally, we would like to end the discussion without suggesting one particular model for the Communities and Crime example, as it not possible to suggest one best solution in the existence of trade-offs. It is ultimately the user who needs to choose a compromise solution which is most suitable for his purposes


\section{Generalization Error}\label{sec:genError}
Until now the focus of the paper has been primarily on minimization of in-sample error. However, the problem with evaluating the models using in-sample error is that the model may demonstrate adequate prediction capabilities on the training sample, but may fail drastically in case of unseen data. Therefore, we are often interested in minimizing the generalization error rather than the in-sample error. In this section, we evaluate MOGA-VS when the in-sample error minimization is replaced with generalization error minimization. We also perform a sample size study showing that for larger sample sizes MOGA-VS is able to produce acceptable results with either of the two estimators.

\subsection{Minimization of Generalization Error}
In order to minimize the generalization error using MOGA-VS, one needs to choose an appropriate generalization error estimator which can be used to compare two models of similar complexity. There are a number of techniques available in the literature which could be used as an estimator for generalization error. In our illustration and experiments we have used the cross validation technique as an estimator to assess the generalization properties of a model.

Once we have chosen an appropriate estimator, it is straightforward to extend the multi-objective approach proposed in this paper for generalization error minimization. If one wishes to minimize the generalization error then the in-sample error minimization objective can be simply replaced with the generalization error minimization objective in MOGA-VS. The generalization error can be estimated using the multi-fold cross validation estimator. Minimizing this error may lead to a set of models that are expected to be less susceptible to over-fitting. In the following steps we briefly describe the estimation of the multi-fold cross validation estimator when the number of folds are chosen to be $k$:
\begin{enumerate}
\item Randomly partition the sample into $k$ equal subsamples
\item Choose one of the subsamples for validation and the remaining $k-1$ subsamples for training the model. Record the MSE value obtained from the validation subsample.
\item Repeat the cross validation task $k$ times, choosing one of the subsamples for validation and the remaining subsamples for training. Take the average of the MSE values obtained from $k$ different validation tasks.
\item The average MSE ($\mbox{MSE}_{cv}$) can be used as an estimator for generalization error.
\end{enumerate}

\begin{figure*}
\begin{minipage}[t]{0.49\linewidth}
\begin{center}
\epsfig{file=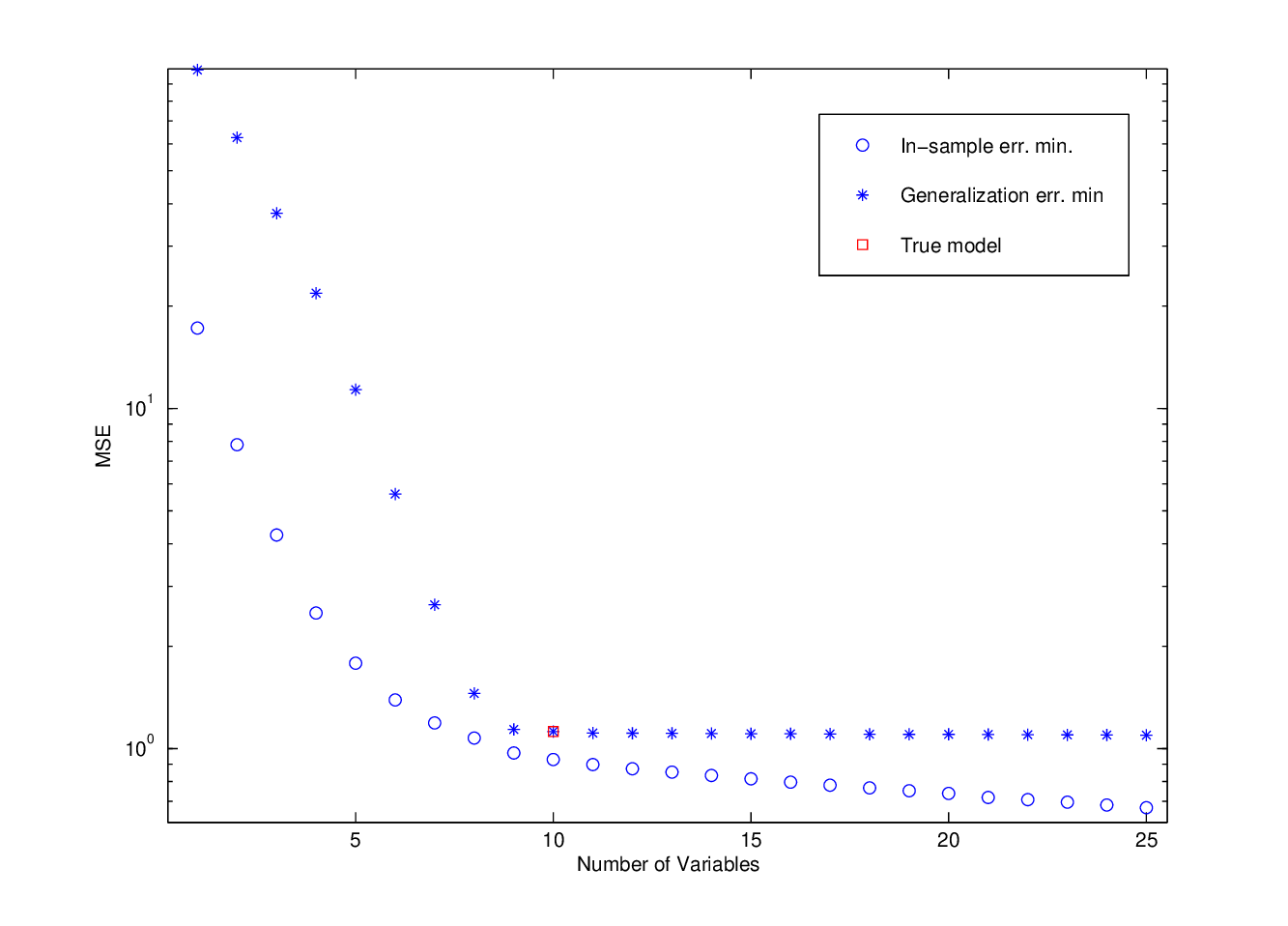,width=0.99\linewidth}
\end{center}
\caption{Models obtained by in-sample error minimization and generalization error minimization using MOGA-VS.}
\label{fig:genError}
\end{minipage}\hfill
\begin{minipage}[t]{0.49\linewidth}
\begin{center}
\epsfig{file=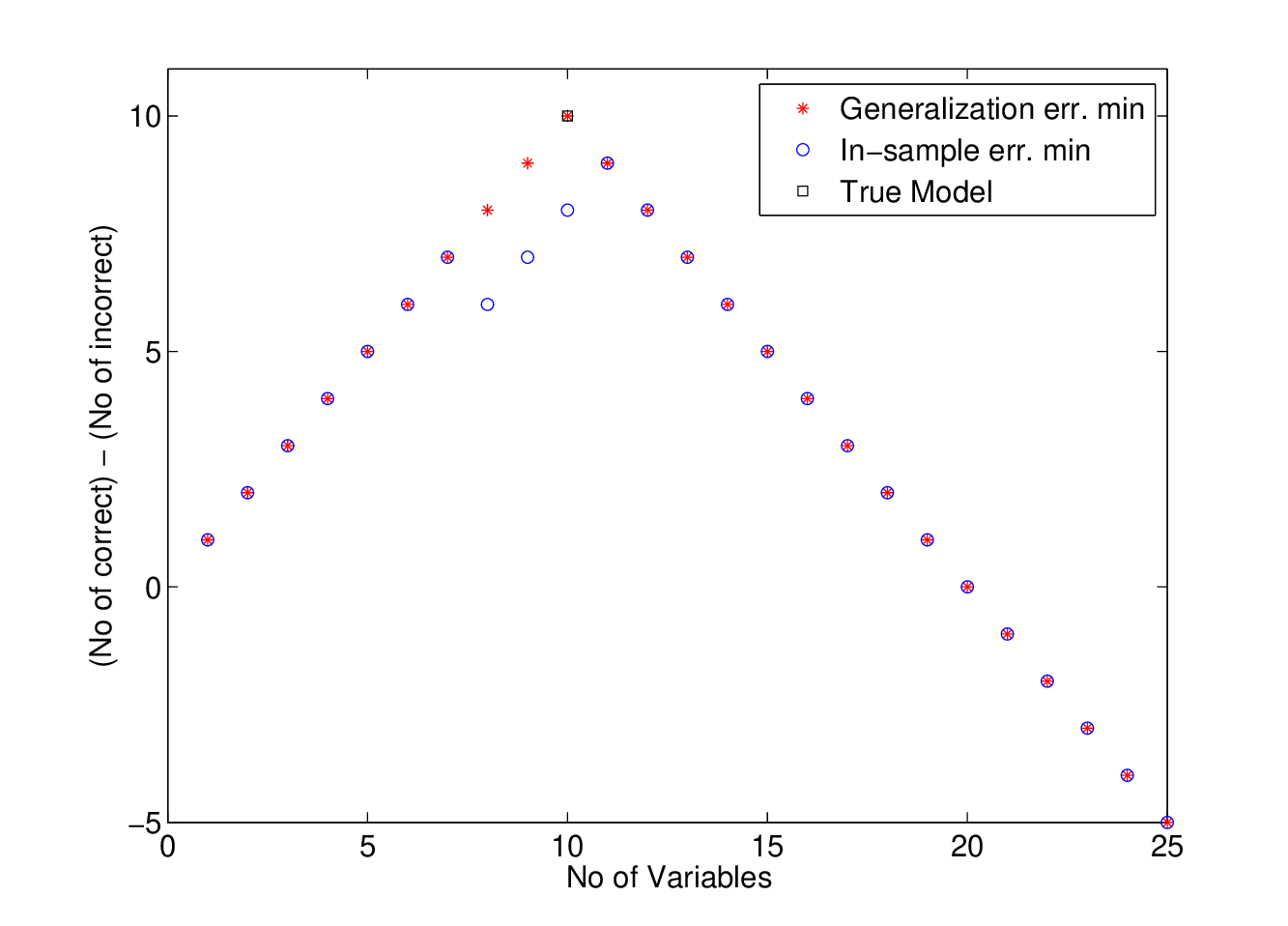,width=0.99\linewidth} 
\end{center}
\caption{The difference between the number of correct and the number of incorrect variables contained in the models produced using generalization error minimization and in-sample error minimization.}
\label{fig:genError2}
\end{minipage}
\end{figure*}

Next, in order to evaluate the idea of minimizing the generalization error obtained using cross validation along with complexity, we execute the MOGA-VS algorithm on simulated example 2 presented in Sub-section \ref{sec:sim2}. The sample size in this experiment is restricted to 200 observations to make the dataset more susceptible to over-fitting. This helps to compare results obtained through in-sample error (MSE) minimization and generalization error ($\mbox{MSE}_{cv}$ obtained from 10-fold cross validation) minimization. 
We execute MOGA-VS algorithm separately for both error minimization criteria to obtain the Pareto-optimal sets of variables corresponding to each complexity level. 

The optimal models for both error minimization measures are presented in Figure \ref{fig:genError}, where y-axis (log scale) represents the MSE values and x-axis represents the complexity. It is clear that the models obtained using in-sample error minimization dominate the true model with respect to the MSE values and complexity as objectives. The most likely explanation for this is over-fitting. 
Further, Figure \ref{fig:genError2} shows the plot of the difference between the number of correct variables and the number of incorrect variables contained in the model. The variables contained in the true model are considered as correct variables and the rest of the variables are considered as incorrect. It can be seen that the frontier corresponding to generalization error minimization contains a model which has all the correct variables and none of the incorrect variables. However, the same is not true for the frontier corresponding to the models obtained using in-sample error minimization. Thus, we observe that for datasets which are more susceptible to over-fitting, it is advisable to perform generalization error minimization instead of in-sample error minimization.

\begin{figure*}
\begin{center}
\epsfig{file=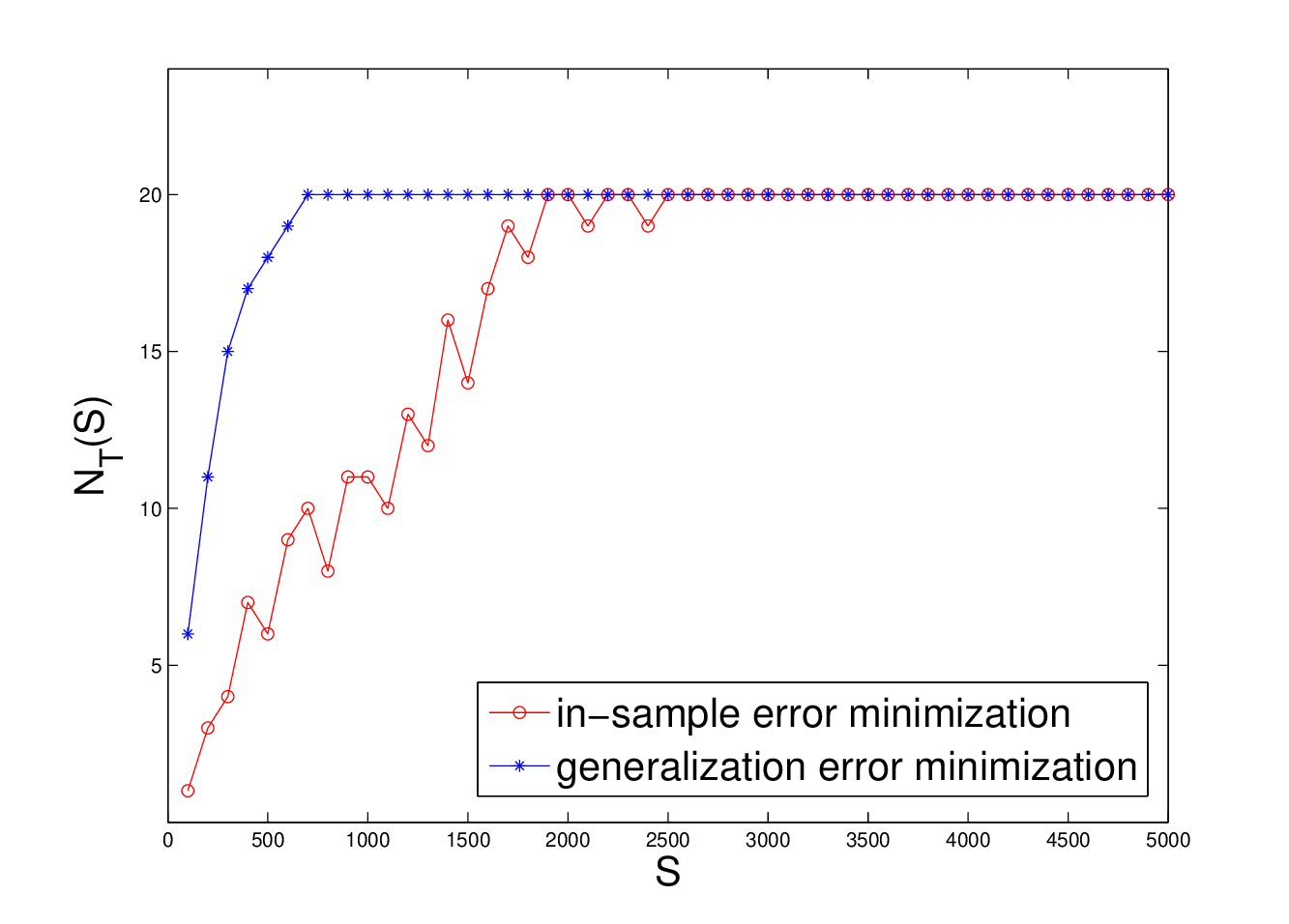,width=0.69\linewidth}
\end{center}
\caption{Number of times MOGA-VS frontier contains the true model ($N_T$) out of 20 datasets of a given sample size ($S$). Results for in-sample error minimization and generalization error minimization using MOGA-VS are shown.}
\label{fig:largeSample}
\end{figure*}

\subsection{Sample size study}
It is a well-known fact that large sample sizes are less susceptible to over-fitting. To evaluate and compare the performance of MOGA-VS on small and large sample sizes, we execute the procedure on simulated example 2 with varying sample sizes. This process is performed for in-sample error minimization as well as generalization error minimization using MOGA-VS.
We vary the sample sizes from 100 to 5000 in steps of 100 and create 20 different datasets for each sample size. MOGA-VS algorithm is then executed on all the datasets. For a given sample size ($S$), we count the number of times ($N_T(S)$) out of 20 runs MOGA-VS frontier contains the true model. The plot for $N_T$ vs $S$ is provided in Figure~\ref{fig:largeSample} for in-sample error minimization as well as generalization error minimization. It can be observed from the figure that with an increase in sample size the MOGA-VS algorithm is able to find the true model most of the times. It is also interesting to note that generalization error minimization finds the true model more often than in-sample error minimization for datasets with smaller sample size. However, the performances are similar for datasets with large sample size. Given that cross-validation techniques can be computationally expensive, we recommend that MOGA-VS may be safely used with in-sample error minimization on datasets with larger sample size.


\section{Conclusions}\label{sec:conclusions}
In this paper, we have proposed a Multi-objective Genetic Algorithm for Variable Selection (MOGA-VS) which can be used for producing the entire set of Pareto-optimal regression models. Once the optimal set of models is known, the most preferred model can be chosen by assessing these models. The proposed algorithm has been tested on a simulated as well as real datasets, and results have been presented. Comparison studies have been performed with state of the art techniques like Lasso, BMA, Stepwise regression and PGA. It has also been shown that the proposed method can be easily used for in-sample as well as out-of-sample error minimization as desired. To conclude, MOGA-VS algorithm can prove to be a useful tool when there are many predictor variables, and a choice for a model with acceptable quality of fit and complexity is to be made. The frontier of solutions produced by the MOGA-VS scheme gives a visual impression to the entire model selection scheme and helps the user to make decisions efficiently. 


\bibliographystyle{natbib}

{}

\end{document}